\documentclass[12pt,titlepage]{article}
\pdfoutput=1

\usepackage[utf8]{inputenc} 
\usepackage{color}
\usepackage{graphicx}
\graphicspath{{figs/}{./}}
\usepackage{tabularx}
\usepackage{amssymb}
\usepackage{amsmath}
\usepackage{authblk}
\usepackage[margin=1in]{geometry}
\usepackage[backend=biber, style=numeric-comp,sorting=none]{biblatex}
\bibliography{apbs}
\usepackage[hidelinks]{hyperref}

\newcommand{\PNNLaffil}{$\star$}
\newcommand{\UCBaffil}{$\dagger$}
\newcommand{\UMaffil}{$\ddagger$}
\newcommand{\SMUaffil}{$\S$}
\newcommand{\SLUaffil}{$\circ$}
\newcommand{\FLaffil}{$\|$}
\newcommand{\Novoaffil}{$\P$}
\newcommand{\UCSDaffil}{$\Delta$}
\newcommand{\PIaffil}{$\lhd$}
\newcommand{\MSUaffil}{$\nabla$}
\newcommand{\Pittaffil}{$\rhd$}
\newcommand{\keyword}[1]{\texttt{#1}}
\newcommand{\param}[1]{$\langle$\textit{#1}$\rangle$}
\newcommand{\revision}[1]{#1}

\begin{document}

\title{Improvements to the APBS biomolecular solvation software suite}

\author[\PNNLaffil]{Elizabeth~Jurrus}
\author[\PNNLaffil]{Dave~Engel}
\author[\PNNLaffil]{Keith~Star}
\author[\PNNLaffil]{Kyle~Monson}
\author[\PNNLaffil]{Juan~Brandi}
\author[\UCBaffil]{Lisa~E.~Felberg}
\author[\UCBaffil]{David~H.~Brookes}
\author[\UMaffil]{Leighton~Wilson}
\author[\SMUaffil]{Jiahui~Chen}
\author[\PNNLaffil]{Karina~Liles}
\author[\PNNLaffil]{Minju~Chun}
\author[\PNNLaffil]{Peter~Li}
\author[\SLUaffil]{David~W.~Gohara}
\author[\FLaffil]{Todd~Dolinsky}
\author[\UCSDaffil]{Robert~Konecny}
\author[\Pittaffil]{David~R.~Koes}
\author[\Novoaffil]{Jens~Erik~Nielsen}
\author[\UCBaffil]{Teresa~Head-Gordon}
\author[\SMUaffil]{Weihua~Geng}
\author[\UMaffil]{Robert~Krasny}
\author[\MSUaffil]{Guo-Wei~Wei}
\author[\UCSDaffil]{Michael~J.~Holst}
\author[\UCSDaffil]{J.~Andrew~McCammon}
\author[\PIaffil]{Nathan~A.~Baker}

\affil[\PNNLaffil]{Pacific Northwest National Laboratory}
\affil[\UCBaffil]{University of California, Berkeley}
\affil[\UMaffil]{University of Michigan}
\affil[\SMUaffil]{Southern Methodist University}
\affil[\SLUaffil]{St.\ Louis University}
\affil[\FLaffil]{FoodLogiQ}
\affil[\Pittaffil]{University of Pittsburgh}
\affil[\Novoaffil]{Protein Engineering, Novozymes A/S}
\affil[\MSUaffil]{Michigan State University}
\affil[\UCSDaffil]{University of California San Diego} 
\affil[\PIaffil]{To whom correspondence should be addressed. Advanced Computing, Mathematics, and Data Division; Pacific Northwest National Laboratory; Richland, WA 99352, USA.  Division of Applied Mathematics; Brown University; Providence, RI 02912, USA. Email:~\href{mailto:nathan.baker@pnnl.gov}{nathan.baker@pnnl.gov}}

\maketitle


\begin{abstract}
The Adaptive Poisson-Boltzmann Solver (APBS) software was developed to solve the equations of continuum electrostatics for large biomolecular assemblages that has provided impact in the study of a broad range of chemical, biological, and biomedical applications.
APBS addresses three key technology challenges for understanding solvation and electrostatics in biomedical applications: accurate and efficient models for biomolecular solvation and electrostatics, robust and scalable software for applying those theories to biomolecular systems, and mechanisms for sharing and analyzing biomolecular electrostatics data in the scientific community.
To address new research applications and advancing computational capabilities, we have continually updated APBS and its suite of accompanying software since its release in 2001.
In this manuscript, we discuss the models and capabilities that have recently been implemented within the APBS software package including: a Poisson-Boltzmann analytical and a semi-analytical solver, an optimized boundary element solver, a geometry-based geometric flow solvation model, a graph theory based algorithm for determining p$K_a$ values, and an improved web-based visualization tool for viewing electrostatics.
\end{abstract}

\section{Introduction}
Robust models of electrostatic interactions are important for understanding early molecular recognition events where long-ranged intermolecular interactions and the effects of solvation on biomolecular processes dominate.
While explicit electrostatic models that treat the solute and solvent in atomic detail are common, these approaches generally require extensive equilibration and sampling to converge properties of interest in the statistical ensemble of interest~\cite{Ren2012}.
\revision{Continuum approaches that integrate out important, but largely uninteresting degrees of freedom, sacrifice numerical precision in favor of robust but qualitative accuracy and efficiency by eliminating the need for sampling and equilibration associated with explicit solute and solvent models.}

While there is a choice among several implicit solvation models~\cite{Davis1990, Perutz1978, Ren2012, Sharp1990, Roux1999, Warshel2006}, one of the most popular implicit solvent models for biomolecules is based on the Poisson-Boltzmann (PB) equation~\cite{Fixman1979, Grochowski2008, Lamm2003}.
The PB equation provides a global solution for the electrostatic potential ($\phi$) within and around a biomolecule by solving the partial differential equation
\begin{equation}
	-\nabla \cdot \epsilon \nabla \phi - \sum_i^M c_i q_i e^{-\beta \left(q_i \phi + V_i \right)} = \rho.
	\label{eqn:pbe}
\end{equation}
The solvent is described by the bulk solvent dielectric constant $\varepsilon_s$ as well as the properties of $i=1,\ldots,M$ mobile ion species, described by their charges $q_i$, concentrations $c_i$, and steric ion-solute interaction potential $V_i$.
The biomolecular structure is incorporated into the equation through $V_i$, a dielectric coefficient function $\epsilon$, and a charge distribution function $\rho$.
The dielectric $\epsilon$ is often set to a constant value $\varepsilon_m$ in the interior of the molecule and varies sharply across the molecular boundary to the value $\varepsilon_s$ which describes the bulk solvent.
The shape of the boundary is determined by the size and location of the solute atoms as well as model-specific parameters such as a characteristic solvent molecule size~\cite{Lee1971}.
The charge distribution $\rho$ is usually a sum of Dirac delta distributions which are located at atom centers.
Finally, $\beta = \left( kT \right)^{-1}$ is the inverse thermal energy where $k$ is the Boltzmann constant and $T$ is the temperature.
The potential $\phi$ can be used in a variety of applications, including visualization, other structural analyses, diffusion simulations, and a number of other calculations that require global electrostatic properties.
The PB theory is approximate and, as a result, has several well-known limitations which can affect its accuracy, particularly for strongly charged systems or high salt concentrations~\cite{Fixman1979, Netz2000}.
Despite these limitations, PB methods are still very important and popular for biomolecular structural analysis, modeling, and simulation. 

\begin{figure}
	\centering
	\includegraphics[width=0.60\linewidth]{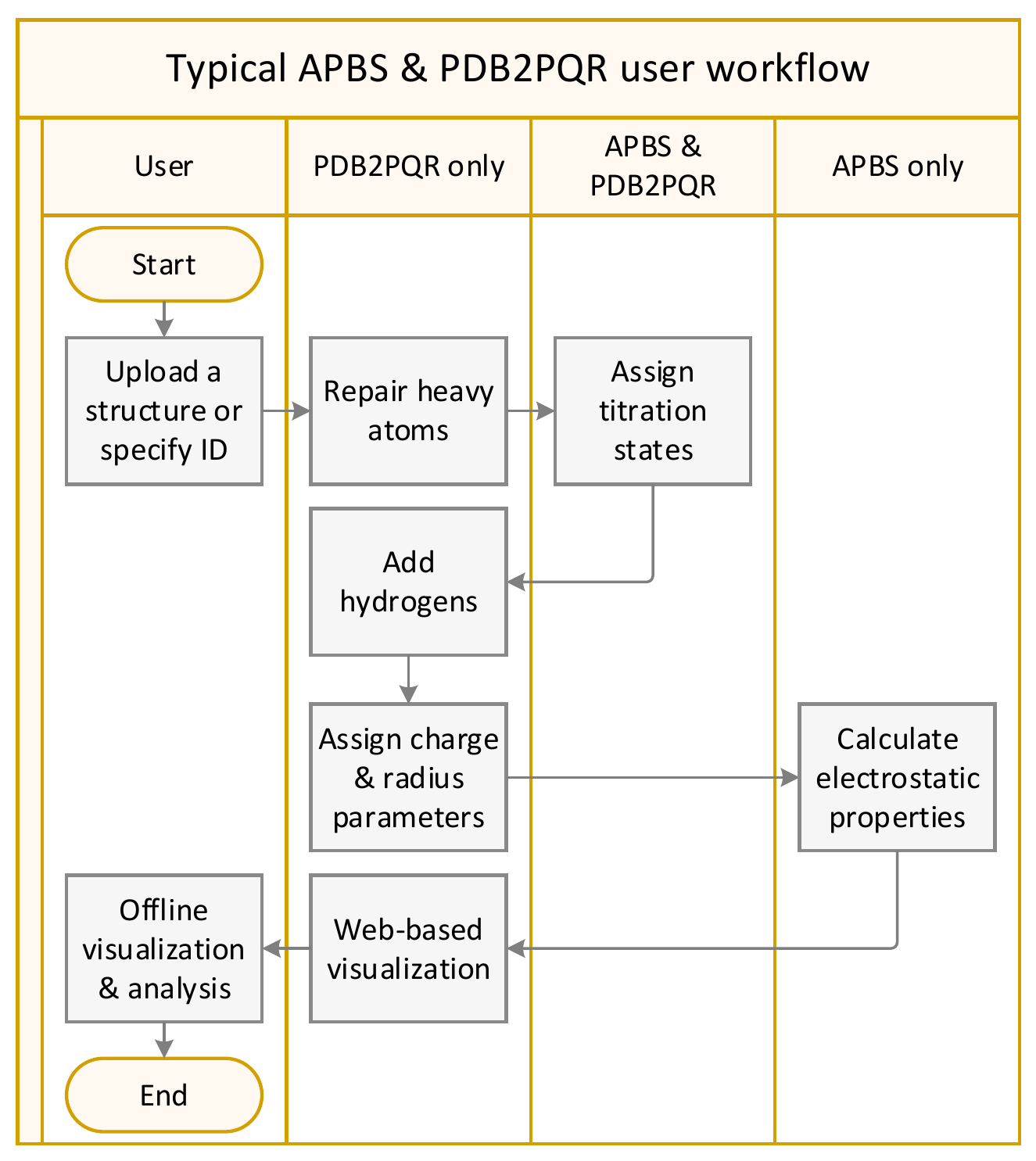} 
	\caption{Workflow for biomolecular electrostatics calculations using the APBS-PDB2PQR software suite. This workflow is supported by the APBS tool suite with specific support by PDB2PQR for preparing biomolecular structures (see Section~\ref{sec:prep}) and support by APBS for performing electrostatics calculations (see Section~\ref{sec:apbs}).}
	\label{fig:PDB2PQR-APBS}
\end{figure}

Several software packages have been developed that solve the Poisson-Boltzmann equations to evaluate energies, potentials, and other solvation properties.
The most significant (based on user base and citations) of these include CHARMM~\cite{Brooks2009}, AMBER~\cite{Case2005}, DelPhi~\cite{Sarkar2012}, Jaguar~\cite{Bochevarov2013}, Zap~\cite{Grant2001}, MIBPB~\cite{Zhou2008}, and APBS~\cite{Baker2001a}.
However the APBS and associated software package PDB2PQR has served a large community of $\sim$27,000 users by creating web servers linked from the APBS website~\cite{APBSweb} that support preparation of biomolecular structures (see Section~\ref{sec:prep}) and a fast finite-difference solution of the Poisson-Boltzmann equation (see Section~\ref{sec:fd}) that are further augmented with a set of analysis tools.
Most APBS electrostatics calculations follow the general workflow outlined in Figure~\ref{fig:PDB2PQR-APBS}.
An even broader range of features and more flexible configuration is available when APBS and PDB2PQR are run from the command line on Linux, Mac, and Windows platforms, and which can be run locally or through web services provided by the NBCR-developed Opal toolkit~\cite{Krishnan2009}.
This toolkit allows for the computing load for processor intensive scientific applications to be shifted to remote computing resources such as those provided by the National Biomedical Computation Resource (NCBR).
Finally, APBS can run through other molecular simulation programs such as AMBER~\cite{Case2005}, CHARMM~\cite{Brooks2009}, NAMD~\cite{Phillips2005}, Rosetta~\cite{Rosetta} and TINKER~\cite{TINKER}.
General support for integration of APBS with 3\textsuperscript{rd}-party programs is provided by the iAPBS library~\cite{Konecny2012, iAPBSweb}.

This article provides an overview of the new capabilities in the APBS software and its associated tools since their release in 2001~\cite{Holst2000, Baker2000, Baker2001a, Baker2001}.
In particular new solutions to the PB equation have been developed and incorporated into APBS: a fully analytical model based on simple spherical geometries that treats full mutual polarization accurately, known as PBAM (Poisson-Boltzmann Analytical Model)~\cite{Lotan2006, Felberg2017} as well as its extension to a semi-analytical PBE solver PB-SAM (Poisson-Boltzmann Semi-Analytical Model) that treats arbitrary shaped dielectric boundaries~\cite{Yap2010, Yap2013}.
APBS-PDB2PQR also now includes an optimized boundary element solver, a geometry-based geometric flow solvation model, a graph theory based algorithm for determining p$K_a$ values, and an improved web-based visualization tool for viewing electrostatics.
We describe these new approaches in the remainder of the paper, and more detailed documentation for these tools is available on the APBS website~\cite{APBSweb}.
\revision{As discussed in the following sections, APBS-PDB2PQR offers a wide range of features for users across all levels of computational biology expertise.}

\section{Preparing biomolecular structures} \label{sec:prep}
Electrostatics calculations begin with specification of the molecule structure and parameters for the charge and size of the constituent atoms.
\revision{Constituent atoms are generally grouped in types with charge and size values specified by atom type in a variety of force field files developed for implicit solvent calculations~\cite{Ren2012}.}
APBS incorporates this information into calculations via the ``PQR'' format. PQR is a file format of unknown origins used by several software packages such as MEAD~\cite{Bashford1997} and AutoDock~\cite{Morris2009}.
The PQR file simply replaces the temperature and occupancy columns of a PDB flat file~\cite{PDBflat} with the per-atom charge ($Q$) and radius ($R$).
There are much more elegant ways to implement the PQR functionality through more modern extensible file formats such as mmCIF~\cite{mmCIF} or PDBML~\cite{PDBML}; however, the simple PDB format is still one of the most widely used formats and, therefore, continued use of the PQR format supports broad compatibility across biomolecular modeling tools and workflows.

The PDB2PQR software is part of the APBS suite that was developed to assist with the conversion of PDB files to PQR format~\cite{Dolinsky2004, Dolinsky2007}.
In particular, PDB2PQR automatically sets up, executes, and optimizes the structure for Poisson-Boltzmann electrostatics calculations, outputting a PQR file that can be used with APBS or other modeling software.
Some of the key steps in PDB2PQR are described below.

\paragraph{Repairing missing heavy atoms}
Within PDB2PQR, the PDB file is examined to see if there are missing heavy (non-hydrogen) atoms.
Missing heavy atoms can be rebuilt using standard amino acid topologies in conjunction with existing atomic coordinates to determine new positions for the missing heavy atoms.
\revision{A \textit{debump} option performs very limited minimization of sidechain $\chi$ angles to reduce steric clashes between rebuilt and existing atoms~\cite{Dolinsky2004}.}

\paragraph{Optimizing titration states}
Amino acid tit\-rat\-ion states are important determinants of bio\-molecular (particularly enzymatic) function and can be used to assess functional activity and identify active sites.
The APBS-PDB2PQR system contains several methods for this analysis.
\begin{itemize}
	\item \textit{Empirical methods.}
	PDB2PQR provides an empirical model (PROPKA~\cite{Sondergaard2011}) that uses a heuristic method to compute p$K_a$ perturbations due to desolvation, hydrogen bonding, and charge-charge interactions.
	PROPKA is included with PDB2PQR.
	The empirical PROPKA method has surprising accuracy for fast evaluation of protein p$K_a$ values~\cite{Li2005}.
	\item \textit{Implicit solvent methods.}
	PDB2PQR also contains two methods for using implicit solvent (Poisson-Boltzmann) models for predicting residue titration states.
	The first method uses Metropolis Monte Carlo to \revision{calculate} titration curves and p$K_a$ values (PDB2PKA); however, sampling issues can be a major problem with Monte Carlo methods when searching over the $\mathcal{O}\left(2^N\right)$ titration states of $N$ titratable residues.
	The second method is a new polynomial-time algorithm for the optimization of discrete states in macromolecular systems~\cite{Purvine2016}.
	\revision{The paper by Purvine et al.~\cite{Purvine2016} describes the performance of the new graph cut method compared to existing approaches.
	While the new deterministic method offers advantages for large systems, the traditional stochastic approach is often sufficient for small to moderate systems.}
	This method transforms interaction energies between titratable groups into a graphical flow network.
	The polynomial-time $\mathcal{O}\left(N^4\right)$ behavior makes it possible to rigorously evaluate titration states for much larger proteins than Monte Carlo methods.
\end{itemize}

\paragraph{Adding missing hydrogens}
The majority of PDB entries do not include hydrogen positions.
Given a titration state assignment, PDB2PQR uses Monte Carlo sampling to position hydrogen atoms and optimize the global hydrogen-bonding network in the structure~\cite{Nielsen2001}.
Newly added hydrogen atoms are checked for steric conflicts and optimized via the debumping procedure discussed above.

\paragraph{Assigning charge and radius parameters} 
Given the tit\-rat\-ion state, atomic charges (for $\rho$) and radii (for $\epsilon$ and $V_i$) are assigned based on the chosen force field.
PDB2PQR currently supports charge/radii force fields from AMBER99~\cite{AMBER99}, CHARMM22~\cite{MacKerell2004}, PARSE~\cite{Sitkoff1994}, PEOE\_PB~\cite{Czodrowski2006}, Swanson et al.~\cite{Swanson2005a}, and Tan et al.~\cite{Tan2006}.
\revision{Many of these force fields only provide parameters for amino acid biomolecules. 
The PEOE approach \cite{Czodrowski2006} provides algorithms to parameterize ligands.
However, we welcome contributions for other biomolecular force fields, particularly for lipids and amino acids.}

\section{Solving the Poisson-Boltzmann and related solvation equations} \label{sec:apbs}
The APBS software was designed from the ground up using modern design principles to ensure its ability to interface with other computational packages and evolve as methods and applications change over time.
APBS input files contain several keywords that are generic with respect to the type of calculation being performed; these are described in Appendix~\ref{app:general}.
The remainder of this section describes the specific solvers available for electrostatic calculations, also described in more detail on the APBS website~\cite{APBSweb}.

\subsection{Finite difference and finite element solvers} \label{sec:fd}
The original version of APBS was based on two key libraries from the Holst research group.
FEtk is a general-purpose multi-level adaptive finite element library~\cite{FEtk, FEtkweb}.
Adaptive finite element methods can resolve extremely fine features of a complex system (like biomolecules) while solving the associated equations over large problem domain.
For example, FEtk has been used to solve electrostatic and diffusion equations over six orders of magnitude in length scale~\cite{Tai2003}.
The finite difference PMG solver~\cite{FEtkweb, Holst1993} trades speed and efficiency for the high-accuracy and high-detail solutions of the finite element FEtk library.
However, many APBS users need only a relatively coarse-grained solution of $\phi$ for their visualization or simulation applications.
Therefore, most APBS users employ the Holst group's finite difference grid-based PMG solver for biomolecular electrostatics calculations.
Appendix~\ref{app:fd} describes some of the common configuration options for finite difference and finite element calculations in APBS.

\subsection{Geometric flow} \label{sec:geoflow}
Several recent papers have described our work on a geometric flow formulation of Poisson-based implicit solvent models~\cite{Chen2010, Chen2011, Chen2012, Daily2013, Thomas2013}.
\revision{The components of this geometric model are described in previous publications \cite{Daily2013, Chen2012}.}
The geometric flow approach couples the polar and nonpolar components of the implicit solvent model with two primary \revision{advantages over existing methods}.
First, this coupling eliminates the need for an \textit{ad hoc} geometric definition for the solute-solvent boundary.
In particular, the solute-solvent interface is optimized as part of the geometric flow calculation.
Second, the optimization of this boundary ensures self-consistent calculation of polar and nonpolar energetic contributions (using the same surface definitions, etc.), thereby reducing confusion and the likelihood of user error.
Additional information about the geometric flow implementation in APBS is provided in Appendix~\ref{app:geoflow}.
This equation is solved in APBS using a finite difference method.

\subsection{Boundary element methods} \label{sec:bem}
\revision{Boundary element methods offer the ability to focus numerical effort on a much smaller region of the problem domain:  the interface between the molecule and the solvent.
APBS now includes a treecode accelerated boundary integral PB solver (TABI-PB) developed by Geng and Krasny to solve a linearized version of the PB equation (Eq.~\ref{eqn:pbe})~\cite{Geng2013}.
A discussion of the relative merits between finite difference/element and boundary integral PB methods are provided in Geng et al.\ and Li et al.~\cite{Geng2013,Li2009}.}
In this method, two coupled integral equations defined on the solute-solvent boundary define a mathematical relationship between the electrostatic surface potential and its normal derivative with a set of integral kernels consisting of Coulomb and screened Coulomb potentials with their normal derivatives~\cite{Juffer1991}.
The boundary element method requires a surface triangulation, generated by a program such as MSMS~\cite{Sanner1995} or NanoShaper~\cite{Decherchi2013}, on which to discretize the integral equations.
A Cartesian particle-cluster treecode is used to compute matrix-vector products and reduce the computational cost of this dense system from $\mathcal{O}(N^2)$ to $\mathcal{O}(N\log N)$ for $N$ points on the discretized molecular surface~\cite{Li2009, Juffer1991}.
A comparison of electrostatic potential with PDB ID 3app from the APBS multi-grid solution method and the new TABI-PB solver are shown in Figure~\ref{fig:surface-mesher}.
Additional information about the boundary element method and its implementation in APBS is provided in Appendix~\ref{app:bem}.
\begin{figure}
	\centering
	\includegraphics[width=0.80\linewidth]{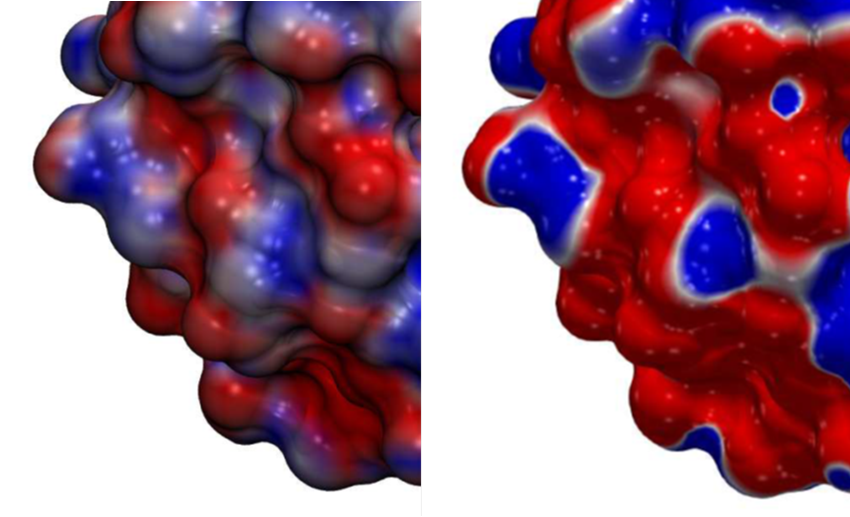} 
	\caption{Electrostatic potential visualization of protein with PDB ID 3app for (A) APBS multigrid and (B) TABI-PB. VMD~\cite{Humphrey1996} was used to generate the figure in (A), and VTK to generate the figure in (B). The potentials are on a $[-4, 4]$ red-white-blue color map in units of kJ/mol/e.
		Calculations were performed at 0.15 M ionic strength in monovalent salt, 298.15 Kelvin, protein dielectric 2, and solvent dielectric 78.
	}
	\label{fig:tabi-mg-comparison}
\end{figure}

\subsection{Analytical and semi-analytical methods} \label{sec:pbam}
\begin{figure}
	\centering
	(A)~\includegraphics[width=0.25\linewidth]{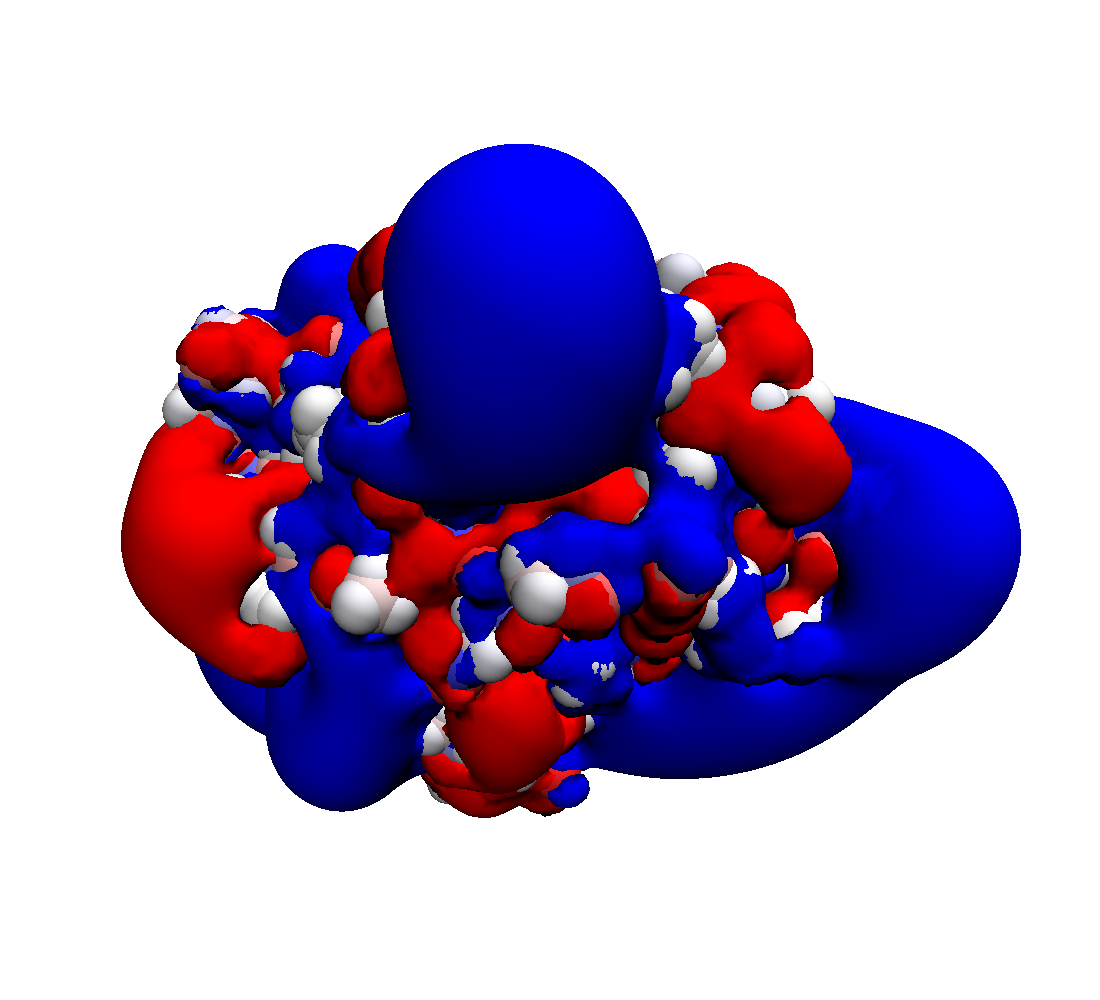} 
	(B)~\includegraphics[width=0.25\linewidth]{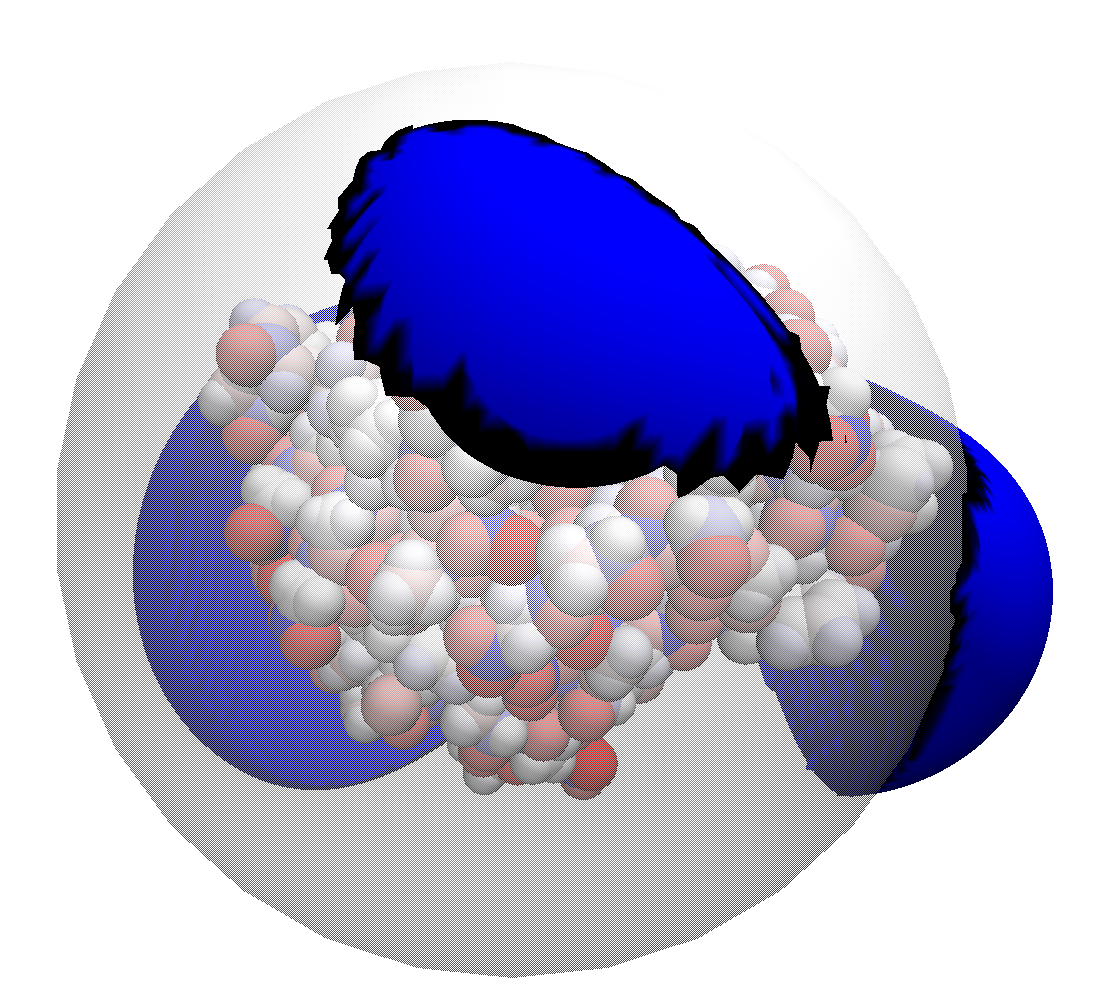} 
	(C)~\includegraphics[width=0.25\linewidth]{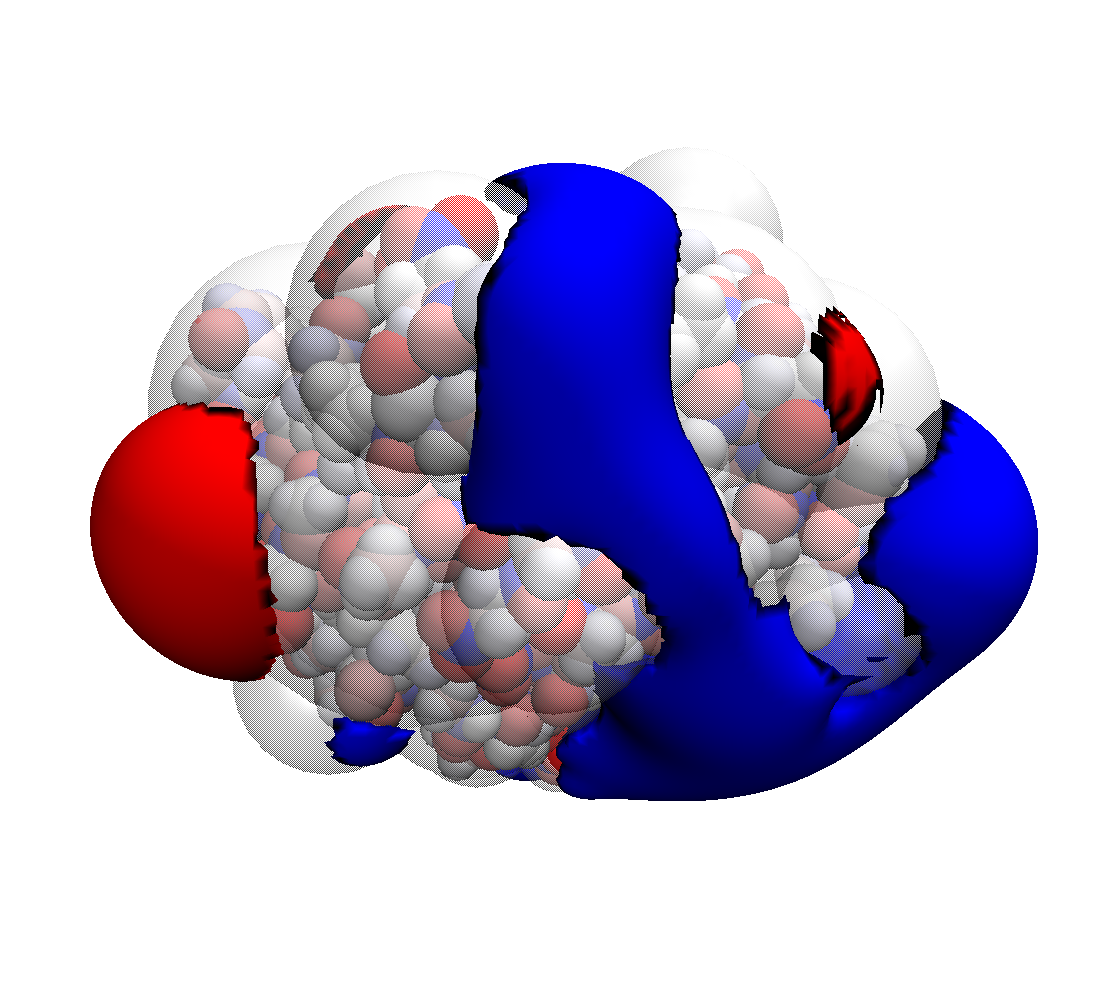}
	(D)~\includegraphics[width=0.23\linewidth]{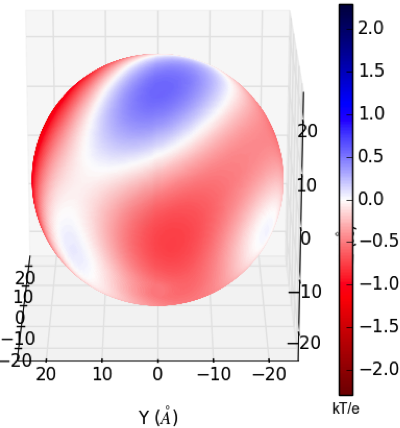} 
	(E)~\includegraphics[width=0.29\linewidth]{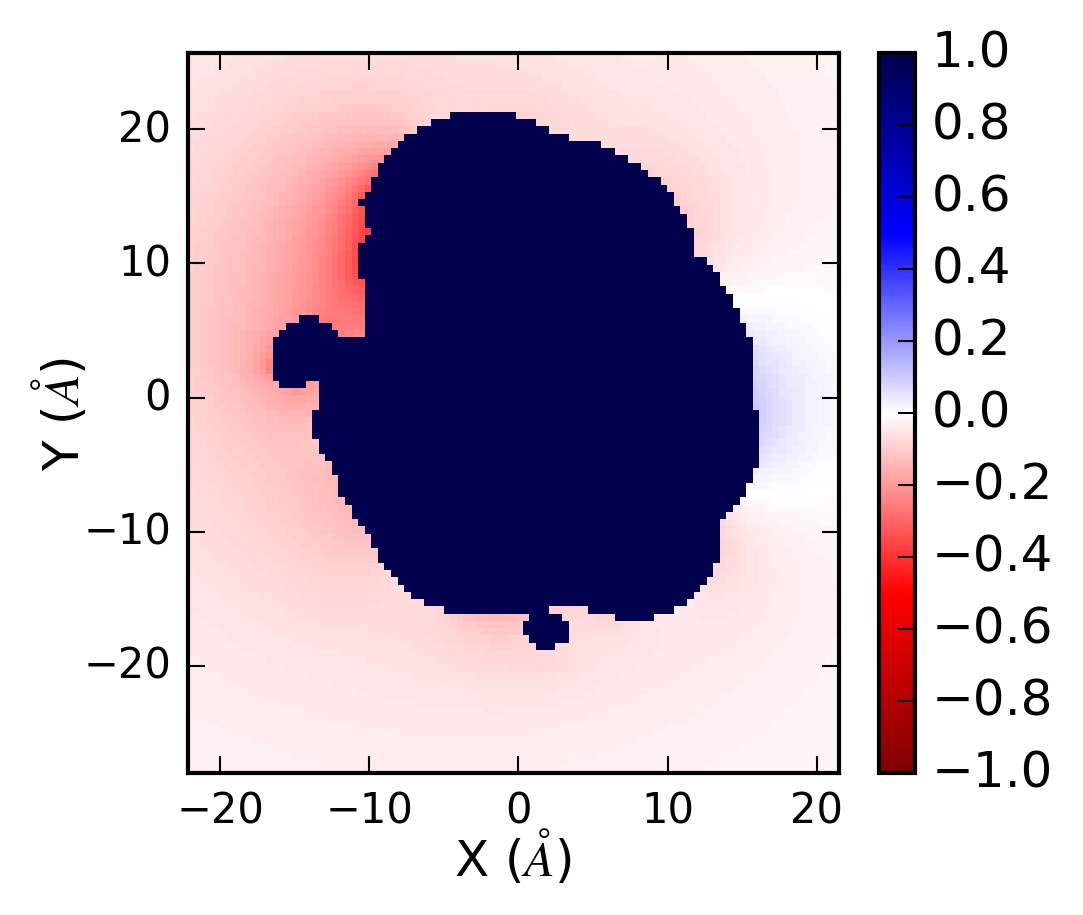} 
	(F)~\includegraphics[width=0.27\linewidth]{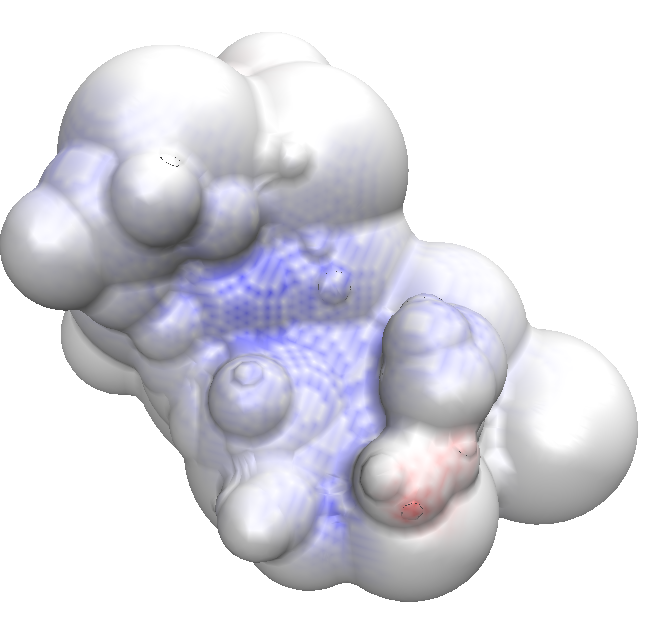}
	\caption{\revision{Comparison of APBS, PB-AM and PB-SAM results (A-C) and electrostatic potential visualization for APBS PB-AM and PB-SAM (D-E).
			VMD~\cite{Humphrey1996} isosurface of barnase molecule generated using
		(A) APBS boundary element method,
		(B) APBS PB-AM method, 
		(C) APBS PB-SAM method. (A-C) atoms colored according to their charge and isosurfaces are drawn at $1.0$ (blue) and $-1.0$ (red) kT/e electrostatic potential.
		(D) APBS PB-AM potential on the coarse-grain surface of the barnase molecule, 
		(E) APBS PB-SAM potential in a 2D plane surrounding the barstar molecule,
		(F) APBS PB-SAM potential over range $[-1, 1]$ on the coarse-grain surface of the barnase molecule (blue region is the location of positive electrostatic potential and barstar association).
		All calculations were performed at 0.0 M ionic strength, 300 Kelvin, pH 7, protein dielectric 2, and solvent dielectric 78.
		All electrostatic potentials are given in units of $kT/e$.}
	}
	\label{fig:pbsam-results}
\end{figure}

Numerical solution methods tend to be computationally intensive, which has led to the adoption \revision{of} analytical approaches for solvation calculations such as generalized Born~\cite{Bashford2000} and the approaches developed by Head-Gordon and implemented in APBS.
In particular, the Poisson-Boltzmann Analytical Method (PB-AM), was developed by Lotan and Head-Gordon in 2006~\cite{Lotan2006}.
PB-AM produces a fully analytical solution to the linearized PB equation for multiple macromolecules, represented as coarse-grained low-dielectric spheres.
This spherical domain enables the use of a multipole expansion to represent charge-charge interactions and higher order cavity polarization effects.
The interactions can then be used to compute physical properties such as interaction energies, forces, and torques.
An example of this approximation, and the resulting electrostatic potentials from PB-AM, are shown in Figure~\ref{fig:pbsam-results}.

The Poisson Boltzmann Semi-Analytical method (PB-SAM) is a modification of PB-AM that incorporates the use of boundary integrals into its formalism to represent a complex molecular domain as a collection of overlapping low dielectric spherical cavities~\cite{Yap2010}.
PB-SAM produces a semi-analytical solution to the linearized PB equation for multiple macromolecules in a screened environment.
This semi-analytical method provides a better representation of the molecular boundary when compared to PB-AM, while maintaining computational efficiency.
An example of this approximation, and the resulting electrostatic potentials from PB-SAM, are shown in Figure~\ref{fig:pbsam-results}.

Because it is fully analytical, PB-AM can be used for model validation as well as for representing systems that are relatively spherical in nature, such as globular proteins and colloids.
PB-SAM, on the other hand has a much more detailed representation of the molecular surface and can therefore be used for many systems that other APBS (numerical) methods are currently used for. Through APBS, both programs can be used to compute the electrostatic potential at any point in space, report energies, forces, and torques of a system of macromolecules, and simulate a system using a BD scheme~\cite{Ermak1978}.
Additional details about these methods and their use in APBS are presented in Appendix~\ref{app:pbam}.
\revision{In addition to these advantages, the strengths and weaknesses of PB-AM and PB-SAM over existing methods are discussion in Yap et al.\ \cite{Yap2010} and Lotan et al.\ \cite{Lotan2006}.}

\section{Using APBS results}
\subsection{Visualization}
\begin{figure} 
	\begin{center}
		\includegraphics[width=.80\textwidth]{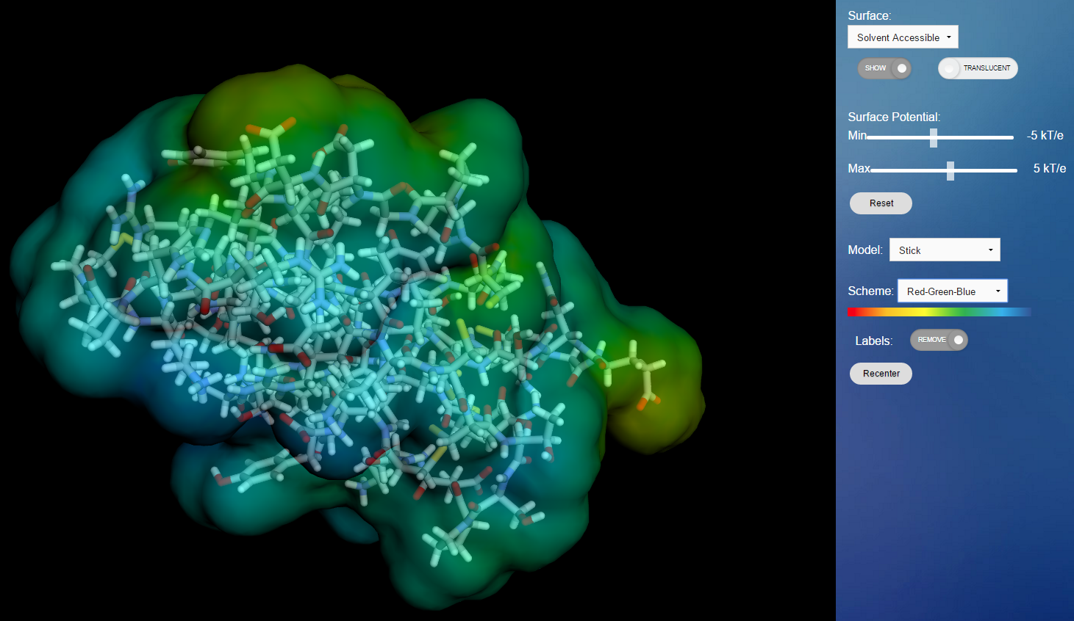}
		\caption{3Dmol.js interface displaying a rendering of fasciculin-2 (1FAS) protein with translucent, solvent accessible surface using a stick model and red-green-blue color scheme. \label{fig:3dmol_interface}}  
	\end{center}  
\end{figure}

\begin{figure}
	\begin{center}
		\includegraphics[width=.75\paperwidth]{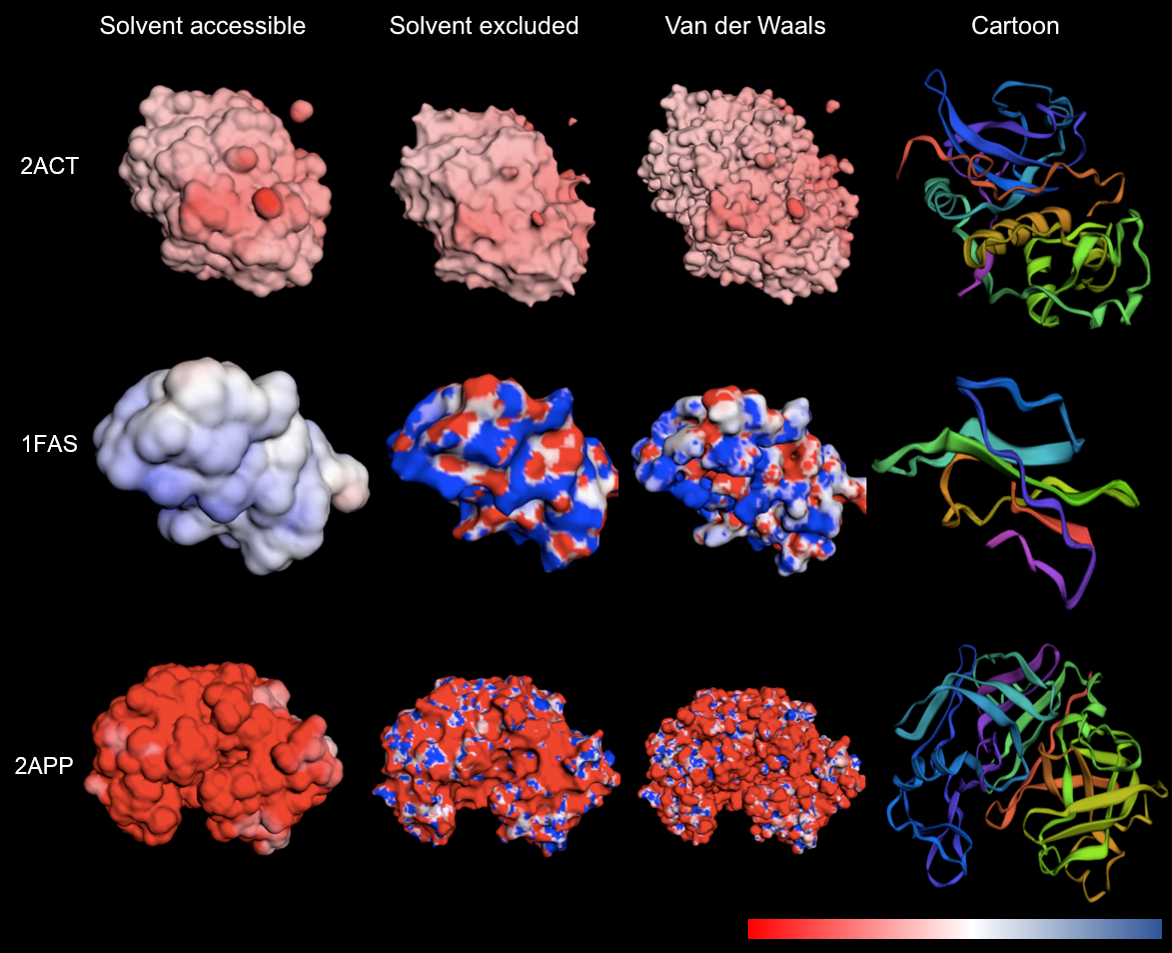}
		\caption{\revision{Renderings of three different proteins: actinidin (2ACT) (top), fasciculin-2 (1FAS) (center), and pepsin-penicillium (2APP) (bottom).}
		To demonstrate the different visualization options.
		From left to right: solvent-accessible surface, solvent-excluded surface, van der Waals  surface, and cartoon models are shown all using red-white-blue color scheme (excluding cartoon model), \revision{where red and blue correspond to negative and positive electrostatic potentials, respectively. \label{fig:features}}}
	\end{center}  
\end{figure}

One of the primary uses of the APBS tools is to generate electrostatic potentials for use in biomolecular visualization software.
These packages offer both the ability to visualize APBS results as well as a graphical interface for setting up the calculation.
Several of these software packages are thick clients that run from users' computers, including PyMOL~\cite{PyMOL, PyMOLweb}, VMD~\cite{VMDweb, Humphrey1996}, PMV~\cite{PMV, MGLToolsweb}, and Chimera~\cite{Chimeraweb, Pettersen2004}.
We have also worked with the developers of Jmol~\cite{Jmolweb, Herraez2006} and 3Dmol.js~\cite{3Dmolweb, Rego2015} to provide web-based setup and visualization of APBS-PDB2PQR calculations and related workflows.
APBS integration with Jmol has been described previously~\cite{Unni2011}.
3Dmol.js is a molecular viewer that offers the performance of a desktop application and convenience of a web-based viewer which broadens accessibility for all users.
As part of the integration with 3Dmol.js, we implemented additional enhancements, including extending our output file formats and creating a customized user interface.
Data from the APBS output file is used to generate surfaces, apply color schemes, and display different molecular styles such as cartoons and spheres.
An example of the 3Dmol.js interface is shown in Figure~\ref{fig:3dmol_interface}.
Examples of 3Dmol.js visualization options are shown in Figure~\ref{fig:features}.

\subsection{Other applications}
Besides visualization and the processes described in Section~\ref{sec:prep}, there are a number of other applications where APBS can be used.
For example, during the past four years, the APBS-PDB2PQR software has been used in the post-simulation energetic analyses of molecular dynamics trajectories~\cite{Dror2013}, understanding protein-nanoparticle interactions~\cite{Treuel2013, DePaoli2014}, understanding nucleic acid-ion interactions~\cite{Lipfert2014}, biomolecular docking~\cite{Roberts2013} and ligand binding~\cite{Evangelidis2009}, developing new coarse-grained protein models~\cite{Spiga2013}, setting up membrane protein simulations~\cite{Stansfeld2015}, etc.
APBS also plays a key role in PIPSA for protein surface electrostatics analysis~\cite{Richter2008} as well as BrownDye~\cite{BrownDye} and SDA~\cite{Martinez2015} for simulation of protein-protein association kinetics through Brownian dynamics.
As discussed above with PB-SAM, another application area for implicit solvent methods is in the evaluation of biomolecular kinetics where implicit solvent models are generally used to provide solvation forces (or energies) for performing (or analyzing) discrete molecular or continuum diffusion simulations with APBS in both of these areas~\cite{Dror2013, Martinez2015, Cheng2007, Cheng2007a, Song2004a, Song2004a, Elcock2004, Mereghetti2012}.

\section{The future of APBS}
\revision{To help with modularity and to facilitate extensibility, APBS was based on an object- oriented programming paradigm with strict ANSI-C compliance.
This ``Clean OO C''~\cite{CleanOOC} has enabled the long-term sustainability of APBS by combining an object-oriented design framework with the portability and speed of C.}
This object-oriented design framework has made it relatively straightforward to extend APBS functionality and incorporate new features.

The Clean OO C design has served APBS well for the past 17 years.
This design still forms the basis for many modules of APBS and PDB2PQR.
Additionally, we have begun to explore a framework for integrating components that exploits the common workflows used by APBS-PDB2PQR users and maps naturally to cloud-based resources.
Our vision for APBS is to build the infrastructure that can enable our users to implement their own models and methods so that they can run on a common system.
Our goal is to have a well-designed, well-tested and well-documented industry-grade framework that will integrate the APBS-PDB2PQR capabilities and make straightforward the incorporation of new features and new workflows.

\section*{Acknowledgments}
The authors gratefully acknowledge NIH grant GM069702 for support of APBS and PDB\-2\-PQR.
PNNL is operated by Battelle for the U.S. DOE under contract DE-AC05-76RL01830.  
THG and DHB were supported by the Director, Office of Science, Office of Basic Energy Sciences, of the U.S.~Department of Energy under contract DE-AC02-05CH11231.
LEF was supported by the National Science Foundation Graduate Research Fellowship under Grant No.~DGE 110640.
JAM and RK were supported by NIH grants GM31749 and GM103426.
LWW was supported by the Department of Defense through the National Defense Science \& Engineering Graduate Fellowship (NDSEG) Program.
WG and RK were supported by NSF grant DMS-1418966.

\pagebreak
\appendix

\section{Appendix}
\renewcommand{\thefigure}{\thesection\arabic{figure}}
\setcounter{figure}{0}

These appendices provide basic information about configuring APBS electrostatics calculations. 
Rather than duplicating the user manual on the APBS website~\cite{APBSweb}, we have only focused on input file keywords that directly relate to the setup and execution of solvation calculations.

\subsection{General keywords for implicit solvent calculations} \label{app:general}
This section contains information about the general keywords used to configure implicit solvent calculations in the \keyword{ELEC} section of APBS input files.
These keywords are used in all of the solver types described in this paper:
\begin{itemize}
	\item \keyword{ion} \keyword{charge} \param{charge} \keyword{conc} \param{conc} \keyword{radius} \param{radius}:  This line can appear multiple times to specify the ionic species in the calculation.
	This specifies the following terms in Eq.~\ref{eqn:pbe}: \param{charge} gives $q_i$ in units of electrons, \param{conc} gives $c_i$ in units of M, and \param{radius} specifies the ionic radius (in \AA) used to calculate $V_i$.
	\item \keyword{lpbe|npbe}: This keyword indicates whether to solve the full nonlinear version of Eq.~\ref{eqn:pbe} (\keyword{npbe}) or a linearized version (\keyword{lpbe}).
	\item \keyword{mol} \param{id}:  Specify the ID of the molecule on which the calculations are to be performed.
	This ID is determined by \keyword{READ} statements which specify the molecules to import.
	\item \keyword{pdie} and \keyword{sdie}:  These keywords specify the di\-elec\-tric coefficient values for the bio\-mol\-ec\-ular interior (\keyword{pdie}) and bulk solvent (\keyword{sdie}).
	A typical value for \keyword{sdie} is 78.54; values for \keyword{pdie} are much more variable and often range from 2 to 40.
	\item \keyword{temp} \param{temp}:  The temperature (in K) for the calculation.  A typical value is 298 K.
\end{itemize}
Additionally, \keyword{READ} statements in APBS input files are used to load molecule information, parameter sets, and finite element meshes.
More detailed information about these and other commands can be found on the APBS website~\cite{APBSweb}.

\subsection{Finite element and finite difference calculations in APBS} \label{app:fd}
The finite difference and finite element methods used by APBS have been described extensively in previous publications~\cite{Baker2000, Holst2000, Holst1993, Baker2001, Baker2001a}; this section focuses on the configuration and use of these methods in APBS.

\subsubsection{Finite difference calculation configuration} \label{sec:fd-config}
APBS users have several ways to invoke the PMG finite difference solver~\cite{Holst1993} capabilities of APBS through keywords in the APBS input file \keyword{ELEC} block.
The different types of finite difference calculations include:
\begin{itemize}
	\item \keyword{mg-manual}: This specifies a manually configured multigrid finite difference calculation in APBS.
	\item \keyword{mg-auto}: This specifies an automatically configured multigrid finite difference calculation in APBS, using focusing~\cite{Gilson1988} to increase the resolution of the calculation in areas of interest.
	\item \keyword{mg-para}:  This specifies a parallel version of the multigrid finite difference calculating, using parallel focusing to increase the resolution of the calculation in areas of interest~\cite{Baker2001a}.
\end{itemize}
These different types of calculations have several keywords described in detail on the APBS website~\cite{APBSweb}.
Some of the most important settings are described below.

\paragraph{Boundary conditions}
Boundary conditions must be imposed on the exterior of the finite difference calculation domain.
In general, biomolecular electrostatics calculations use Dirichlet boundary conditions, setting the value of the potential to an asymptotically correct approximation of the true solution.
There are several forms of these boundary conditions available in APBS with approximately equal accuracy given a sufficiently large calculation domain (see below).
The only boundary condition which is \emph{not} recommended for typical calculations is the zero-potential Dirichlet condition.

\paragraph{Grid dimensions, center, and spacing}
Finite difference calculations in APBS are performed in rectangular domains.  
The key aspects of this domain include its length $L_i$ and number of grid points $n_i$ in each direction $i$.
These parameters are related to the spacing $h_i$ of the finite difference grid by $h_i = L_i/(n_i-1)$.
Grid spacings below 0.5 \AA\ are recommended for most calculations.
The number of grid points is specified by the \keyword{dime} keyword.
For \keyword{mg-manual} calculations, the grid spacing can be specified by \emph{either} the \keyword{grid} or the \keyword{glen} keywords, which specify the grid spacing or length, respectively.
For \keyword{mg-auto} or \keyword{mg-para} calculations, the grid spacing is determined by the \keyword{cglen} and \keyword{fglen} keywords, which specify the lengths of the coarse and fine grids for the focusing calculations.
Grid lengths should extend sufficient distance away from the biomolecule so that the chosen boundary condition is accurate. 
In general, setting the length of the coarsest grid to approximately 1.5 times the size of the biomolecule gives reasonable results.
However, as a best practice, it is important to ensure that the calculated quantities of interest do not change significantly with grid spacing or grid length.
The center of the finite difference grid can be specified by the \keyword{gcent} command for \keyword{mg-manual} calculations or by the \keyword{cgcent} and \keyword{fgcent} keywords for the coarse and fine grids (respectively) in \keyword{mg-auto} or \keyword{mg-para} focusing calculations.
These keywords can be used to specify absolute grid centers (in Cartesian coordinates) or relative centers based on molecule location.
Because of errors associated with charge discretization, it is generally a good idea to keep the positions of molecules on a grid fixed for all calculations.
For example, a binding calculation for rigid molecules should keep all molecules in the same positions on the grid.

\paragraph{Charge discretization}
As mentioned above, atomic charge distributions are often modeled as a collection of Dirac delta functions or other basis functions with extremely small spatial support.
However, finite difference calculations \revision{performed} on grids with finite spacing, requiring charges to be mapped across several grid points.
This mapping creates significant dependence of the electrostatic potential on the grid spacing which is why it is always important to use the same grid setup for all parts of an electrostatic calculation.
The \keyword{chgm} keyword controls the interpolation scheme used for charge distributions and includes the following types of discretization schemes:
\begin{itemize}
	\item \keyword{spl0}: Traditional trilinear interpolation (linear splines).
	The charge is mapped onto the nearest-neighbor grid points.
	Resulting potentials are very sensitive to grid spacing, length, and position.
	\item \keyword{spl2}: Cubic B-spline discretization as described by Im et al.~\cite{IBR98}.
	The charge is mapped onto the nearest- and next-nearest-neighbor grid points.
	Resulting potentials are somewhat less sensitive (than \keyword{spl0}) to grid spacing, length, and position; however, this discretization can often require smaller grid spacings for accurate representation of charge positions.
	\item \keyword{spl4}: Quintic B-spline discretization as described by Schnieders et al.~\cite{Schnieders2007}.
	Similar to \keyword{spl2}, except the charge/multipole is additionally mapped to include next-next-nearest neighbors (125 grid points receive charge density).
	This discretization results in less sensitivity to grid spacing and position but nearly always requires smaller grid spacings for accurate representation of charge positions.
\end{itemize}

\paragraph{Surface definition}
APBS provides several difference surface definitions through the \keyword{srfm} keyword:
\begin{itemize}
	\item \keyword{mol}: The dielectric coefficient is defined based on a molecular surface definition.
	The problem domain is divided into two spaces.
	The ``free volume'' space is defined by the union of solvent-sized spheres (size determined by the \keyword{srad} keyword) which do not overlap with biomolecular atoms.
	This free volume is assigned bulk solvent dielectric values.
	The complement of this space is assigned biomolecular dielectric values.
	With a non-zero solvent radius (\keyword{srad}), this choice of coefficient corresponds to the traditional definition used for PB calculations.
	When the solvent radius is set to zero, this corresponds to a van der Waals surface definition.
	The ion-accessibility coefficient is defined by an ``inflated'' van der Waals model.
	Specifically, the radius of each biomolecular atom is increased by the radius of the ion species (as specified with the \keyword{ion} keyword).
	The problem domain is then divided into two spaces.
	The space inside the union of these inflated atomic spheres is assigned an ion-accessibility value of 0; the complement space is assigned bulk ion accessibility values.
	\item \keyword{smol}: The dielectric and ion-accessibility coefficients are defined as for \keyword{mol} (see above).
	However, they are then ``smoothed'' by a 9-point harmonic averaging to somewhat reduce sensitivity to the grid setup~\cite{Bruccoleri1997}.
	\item \keyword{spl2}: The dielectric and ion-accessibility coefficients are defined by a cubic-spline surface~\cite{IBR98}.
	The width of the dielectric interface is controlled by the \keyword{swin} parameter.
	These spline-based surface definitions are very stable with respect to grid parameters and therefore ideal for calculating forces. However, they require substantial reparameterization of the force field~\cite{Nina1999}.
	\item \keyword{spl4}: The dielectric and ion-accessibility coefficients are defined by a 7\textsuperscript{th}-order polynomial.
	This surface definition has characteristics similar to \keyword{spl2}, but provides higher order continuity necessary for stable force calculations with atomic multipole force fields (up to quadrupole)~\cite{Schnieders2007}.
\end{itemize}

\subsubsection{Finite elements calculation configuration}
Users invoke the FEtk finite element solver~\cite{FEtk} in APBS by including the \keyword{fe-manual} keyword in the \keyword{ELEC} section of the input file.
Many aspects of the finite element configuration closely follow the finite difference options described above.
This section only describes the configuration options which are unique to finite element calculations in APBS.

Finite element calculations begin with an initial mesh.
This mesh can be imported from an external file via the \keyword{usemesh} keyword or generated by APBS.
APBS generates the initial mesh from a very coarse 8-tetrahedron mesh of size \keyword{domainLength} which is then refined uniformly or selectively at the molecular surface and charge locations, based on the value of the \keyword{akeyPRE} keyword.  
This initial refinement occurs until the mesh has \keyword{targetNum} vertices or has reached \keyword{targetRes} edge length (in \AA).

As described previously~\cite{Holst2000, Baker2000}, APBS uses FEtk in a solve-estimate-refine iteration which involves the following steps:
\begin{enumerate}
	\item Solve the problem with the current finite element mesh.
	\item Estimate the error in the solution as a function of position on the mesh.
	The method of error estimation is determined by the \keyword{ekey} keyword which can have the values:
	\begin{itemize}
		\item \keyword{simp}: Per-simplex error threshold; simplices with error above this limit are flagged for refinement.
		\item \keyword{global}: Global (whole domain) error limit; flag enough simplices for refinement to reduce the global error below this limit.
		\item \keyword{frac}: The specified fraction of the simplices with the highest amount of error are flagged for refinement.
	\end{itemize}
	\item Adaptively refine the mesh to reduce the error using the error metric described by \keyword{ekey}.
\end{enumerate}
This iteration is repeated until a target error level \keyword{etol} is reached or a maximum number of solve-estimate refine iterations (\keyword{maxsolve}) or vertices (\keyword{maxvert}) is reached.

\subsection{Geometric flow calculations in APBS} \label{app:geoflow}
This section contains additional information about the geometric flow equation implementation in APBS introduced in Section~\ref{sec:geoflow}.
The geometric flow methods used by APBS have been described extensively in previous publications~\cite{Chen2010, Chen2011, Chen2012, Daily2013, Thomas2013a}; this section focuses on the configuration and use of these methods in APBS.

\subsubsection{Geometric flow calculation configuration}
Users invoke the geometric flow solver in APBS by including the \keyword{geoflow-auto} keyword in the \keyword{ELEC} section of the input file.
Because the geometric flow solver is based on finite difference solvers, many of the keywords for this section are similar to those described in Section~\ref{sec:fd-config}.
Three additional parameters are needed for geometric flow calculations to specify how nonpolar solvation is linked to the polar implicit solvent models:
\begin{itemize}
	\item \keyword{gamma} \param{tension}:  Specify the surface tension of the solvent in units of kJ mol$^{-1}$ \AA$^{-2}$.
	Based on Daily et al.~\cite{Daily2013}, a recommended value for small molecules is 0.431 kJ mol$^{-1}$ \AA$^{-2}$.
	\item \keyword{press} \param{pressure}:  Specify the internal pressure of the solvent in units of kJ mol$^{-1}$ \AA$^{-3}$.
	Based on Daily et al.~\cite{Daily2013}, a recommended value for small molecules is 0.104 kJ mol$^{-1}$ \AA$^{-3}$.
	\item \keyword{bconc} \param{concentration}:  Specify the bulk concentration of solvent in \AA$^{-3}$.
	The bulk density of water, 0.0334~\AA$^{-3}$, is recommended.
	\item \keyword{vdwdisp} \param{bool}:  Indicate whether van der Waals interactions should be included in the geometric flow calculation through \param{bool} (1~=~include, 0~=~exclude).
	If these interactions are included, then a force field with van der Waals terms must be included through a \keyword{READ} statement in the APBS input file.
\end{itemize}

\subsection{Boundary element method implementation} \label{app:bem}
This appendix provides additional information about the boundary element method introduced in Section~\ref{sec:bem}.

\subsubsection{Boundary element method background}
\begin{figure}
	\centering
	\hspace{-5pt}
	(A)~\hspace{-3pt}\includegraphics[width=0.25\linewidth]{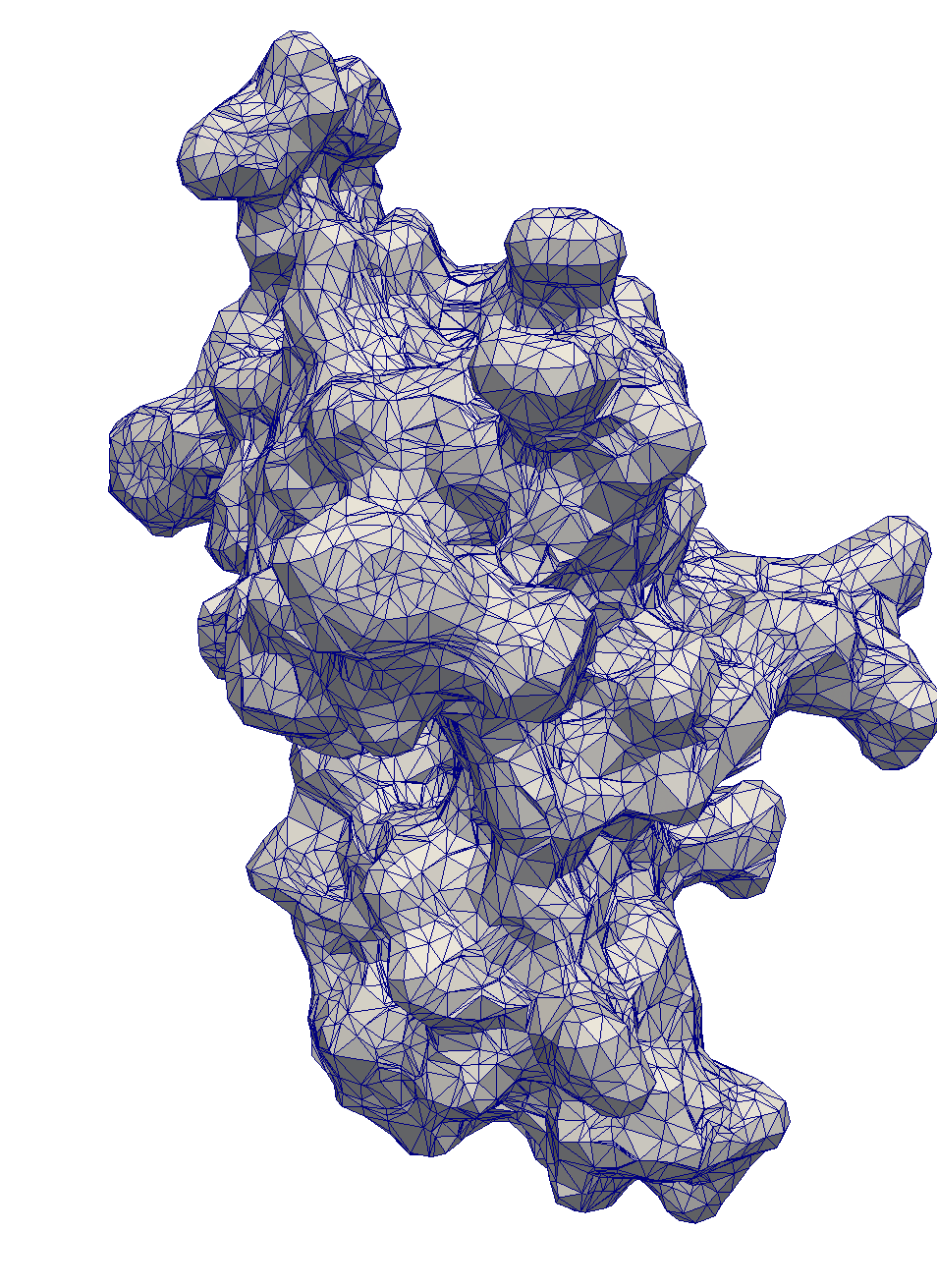} 
	\hspace{-10pt}
	(B)~\hspace{-3pt}\includegraphics[width=0.25\linewidth]{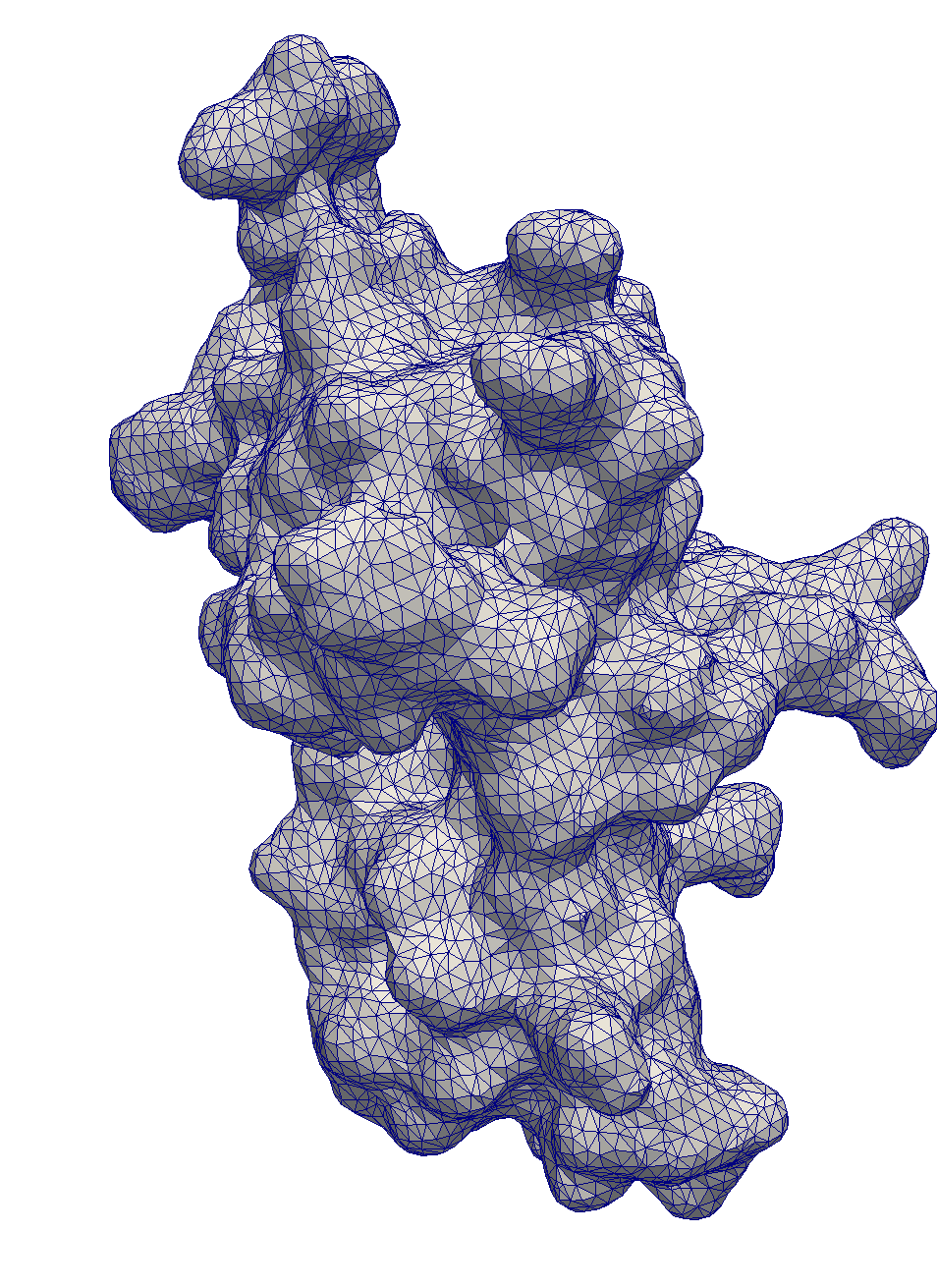}
	\hspace{-10pt} 
	(C)~\hspace{-3pt}\includegraphics[width=0.25\linewidth]{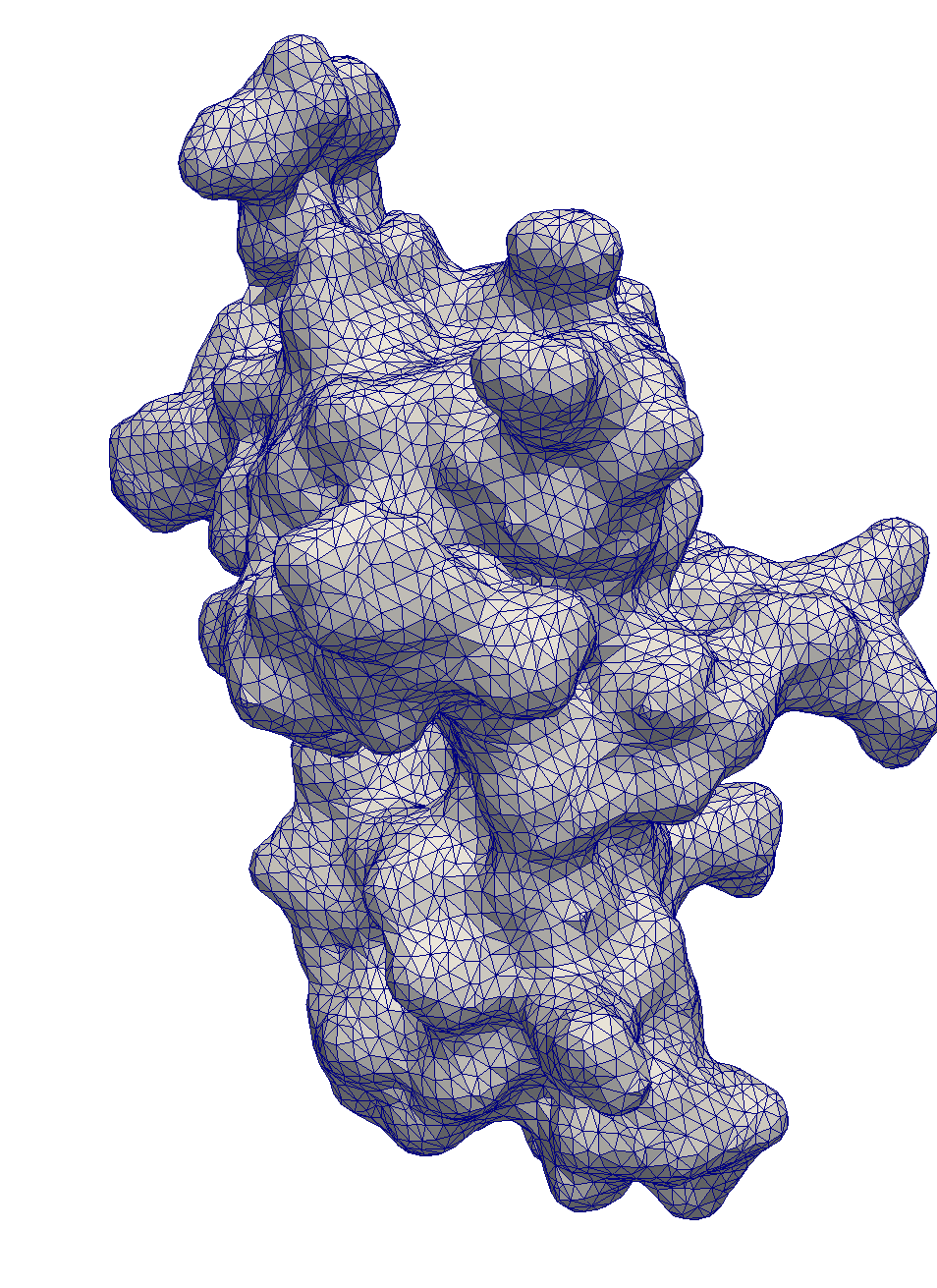}
	\hspace{-5pt}
	(D)~\hspace{-3pt}\includegraphics[angle=0,width=0.25\linewidth]{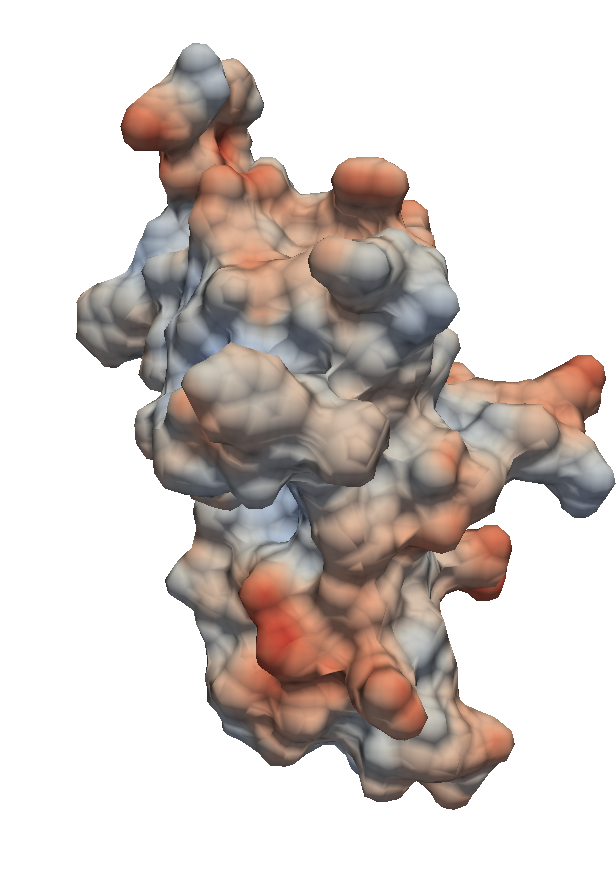} 
	\hspace{-10pt}
	(E)~\hspace{-3pt}\includegraphics[angle=0,width=0.25\linewidth]{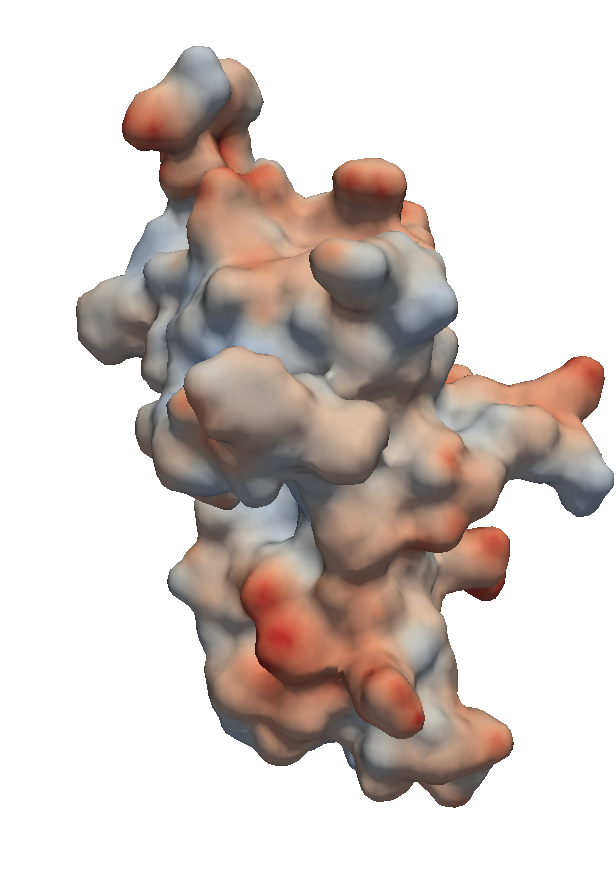}
	\hspace{-10pt} 
	(F)~\hspace{-3pt}\includegraphics[angle=0,width=0.25\linewidth]{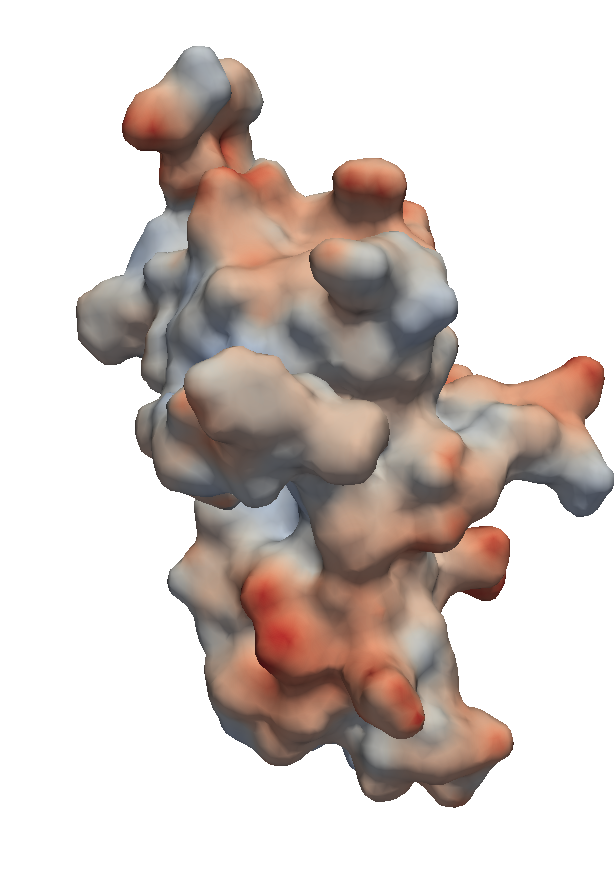}
	\hspace{-5pt}
	\caption{APBS TABI-PB electrostatic potential results for PDB ID 1a63.
		Surfaces were generated with (A) 20228 triangles via MSMS~\cite{Sanner1995}, (B) 20744 triangles via NanoShaper SES, and (C) 21084 triangles via NanoShaper Skin~\cite{Decherchi2013}.
		These discretizations resulted in surface potentials (with units $kT/e$) of (D)~MSMS in range $[-8.7, 8.6]$, (E)~NanoShaper SES in range $[-13.4, 7.5]$, and (F)~NanoShaper Skin in range $[-33.8, 8.0]$.
		All calculations were performed at 0.15 M ionic strength in 1:1 salt, with protein dielectric 1, solvent dielectric 80, and temperature 300K.
		\revision{Red and blue surface colors correspond to negative and positive electrostatic surface potentials, respectively.}
		\label{fig:surface-potential}
		\label{fig:surface-mesher}}
\end{figure}

This section provides additional background on the TABI-PB boundary element solver \cite{Geng2013}, introduced in Section~\ref{sec:bem}.
As described earlier, this method involves solving two coupled integral equations defined on the solute-solvent boundary \revision{which} define a mathematical relationship between the electrostatic surface potential and its normal derivative with a set of integral kernels consisting of Coulomb and screened Coulomb potentials with their normal derivatives.
The boundary element method requires a surface triangulation, generated by a program such as MSMS~\cite{Sanner1995} or NanoShaper~\cite{Decherchi2013}, on which to discretize the integral equations.
Figure~\ref{fig:surface-mesher} shows different types of surface discretizations as well as example electrostatic potential output.

The coupled second kind integral equations employed by TABI-PB for calculating the surface potential $\phi$ and its normal derivative~\cite{Juffer1991} are: 
\begin{equation} \label{eq:integraleq}
	\begin{split}
		\frac{1}{2}\left(1+\varepsilon\right)\phi\left({\bf x}\right) =& \int_\Gamma\left[K_1\left({\bf x}, {\bf y}\right) \frac{\partial \phi\left({\bf y}\right)}{\partial \nu} + K_2\left({\bf x}, {\bf y}\right) \phi\left({\bf y}\right) \right]dS_{\bf y} + S_1({\bf x}), \ {\bf x} \in \Gamma, \\ %
		\frac{1}{2}\left(1+\frac{1}{\varepsilon}\right) \frac{\partial \phi\left({\bf x}\right)}{\partial \nu} =& \int_\Gamma\left[K_3\left({\bf x}, {\bf y}\right) \frac{\partial \phi\left({\bf y}\right)}{\partial \nu} + K_4\left({\bf x}, {\bf y}\right) \phi\left({\bf y}\right) \right]dS_{\bf y} + S_2({\bf x}), \ {\bf x} \in \Gamma 
	\end{split}
\end{equation} 
where $\varepsilon = \varepsilon_m / \varepsilon_s$, the ratio of the dielectric constant in the solute region and the dielectric constant in the solvent region.
The integral kernels $K_1, K_2, K_3, K_4$ are defined in Eq.~\ref{eq:kernels}.
\begin{equation}
	\begin{split}
		K_1\left({\bf x}, {\bf y}\right) &= G_0\left({\bf x}, {\bf y}\right) - G_\kappa\left({\bf x}, {\bf y}\right), \\
		K_2\left({\bf x}, {\bf y}\right) &= \varepsilon \frac{\partial G_\kappa\left({\bf x}, {\bf y}\right)}{\partial \nu_{\bf y}} -\frac{\partial G_0\left({\bf x}, {\bf y}\right)}{\partial \nu_{\bf y}}, \\
		K_3\left({\bf x}, {\bf y}\right) &= \varepsilon \frac{\partial G_0\left({\bf x}, {\bf y}\right)}{\partial \nu_{\bf x}} -\frac{1}{\varepsilon}\frac{\partial G_\kappa\left({\bf x}, {\bf y}\right)}{\partial \nu_{\bf x}}, \\
		K_4\left({\bf x}, {\bf y}\right) &= \varepsilon \frac{\partial^2 G_\kappa\left({\bf x}, {\bf y}\right)}{\partial \nu_{\bf x} \partial \nu_{\bf y}} -\frac{1}{\varepsilon}\frac{\partial^2 G_0\left({\bf x}, {\bf y}\right)}{\partial \nu_{\bf x} \partial \nu_{\bf x}},
		\label{eq:kernels}
	\end{split}
\end{equation}
where $G_0$ and $G_\kappa$ are the Coulomb and screened Coulomb potentials defined as:
\begin{equation}
	\begin{split}
		G_0\left({\bf x}, {\bf y}\right) &= \frac{1}{4\pi \left|{\bf x} - {\bf y}\right|}, \\ G_\kappa\left({\bf x}, {\bf y}\right) &= \frac{e^{-\kappa\left|{\bf x} - {\bf y}\right|}}{4\pi \left|{\bf x} - {\bf y}\right|}.
	\end{split}
\end{equation}
The normal derivatives of the potential kernels $G$ are:
\begin{equation}
	\begin{split}
		\frac{\partial G\left({\bf x}, {\bf y}\right)}{\partial \nu_{\bf y}} =& \sum\limits_{n=1}^{3} \nu_n\left({\bf y}\right)\partial_{y_n}G\left({\bf x}, {\bf y}\right), \\
		\frac{\partial G\left({\bf x}, {\bf y}\right)}{\partial \nu_{\bf x}} =& -\sum\limits_{n=1}^{3} \nu_n\left({\bf y}\right)\partial_{x_n}G\left({\bf x}, {\bf y}\right), \\
		\frac{\partial^2 G\left({\bf x}, {\bf y}\right)}{\partial \nu_{\bf x}\partial \nu_{\bf y}} =& -\sum\limits_{m=1}^{3}\sum\limits_{n=1}^{3} \nu_m\left({\bf x}\right)\nu_n\left({\bf y}\right)\partial_{x_n}\partial_{y_n}G\left({\bf x}, {\bf y}\right) 
	\end{split}
\end{equation}
for the three spatial components $n$ of the normal direction.
Additionally, the source terms $S_1$ and $S_2$ in Eq.~\ref{eq:integraleq} are:
\begin{equation}
	\begin{split}
		S_1({\bf x}) &= \frac{1}{\varepsilon_m}\sum\limits_{k=1}^{N_c}q_k G_0\left({\bf x}, {\bf y}_k\right), \\
		S_2({\bf x}) &= \frac{1}{\varepsilon_m}\sum\limits_{k=1}^{N_c}q_k \frac{\partial G_0\left({\bf x}, {\bf y}_k\right)}{\partial \nu_{\bf x}},
	\end{split}
\end{equation}
where $N_c$ is the number of atoms in the solute molecule, and $q_k$ is the charge of the $k$th atom.
Note that $S_1$ is a linear superposition of the point charge electrostatic potentials, and $S_2$ is a linear superposition of the normal derivatives of the potentials.

Given a surface triangularization with $N$ elements -- where ${\bf x}_i$ and $A_i$ are the centroid and area, respectively, of the $i$th triangle -- the integral equations are discretized as:
\begin{equation}
	\begin{split}
		\frac{1}{2}\left(1+\varepsilon\right)\phi\left({\bf x}_i\right) =& \sum_{\substack{j=1 \\ j\neq i}}^N \left[K_1\left({\bf x}_i, {\bf x}_j\right) \frac{\partial \phi\left({\bf x}_j\right)}{\partial \nu} + K_2\left({\bf x}_i, {\bf x}_j\right) \phi\left({\bf x}_j\right) \right]A_j + S_1({\bf x}_i), \\ %
		\frac{1}{2}\left(1+\frac{1}{\varepsilon}\right) \frac{\partial \phi\left({\bf x}_i\right)}{\partial \nu} =& \sum_{\substack{j=1 \\ j\neq i}}^N \left[K_3\left({\bf x}_i, {\bf x}_j\right) \frac{\partial \phi\left({\bf x}_j\right)}{\partial \nu} + K_4\left({\bf x}_i, {\bf x}_j\right) \phi\left({\bf x}_j\right) \right]A_J + S_2({\bf x}_I).
	\end{split}
\end{equation}
The omission of the $j=i$ term in the summation avoids the singularity of the kernels at that point.
Note that the right hand sides of these equations consist of sums of products of kernels and the surface potential or its normal derivative.
These are the analogues to the $N$-body potential in the treecode, with the surface potential or its normal derivatives playing the role of the charges.
In the discretized form, the total electrostatic energy of solvation is given by Eq.~\ref{eq:tabi_TEE}:
\begin{equation}
	E_\text{sol}=\frac{1}{2}\sum_{k=1}^{N_c}q_k\sum_{j=1}^N \left[K_1\left({\bf x}_k, {\bf x}_j\right) \frac{\partial \phi\left({\bf x}_j\right)}{\partial \nu} + K_2\left({\bf x}_k, {\bf x}_j\right) \phi\left({\bf x}_j\right) \right]A_j
	\label{eq:tabi_TEE}
\end{equation}
where $q_k$ is the charge on the $k$th atom of the solute molecule, and ${\bf x}_k$ is its position.

The matrix-vector products involve evaluation of the integral kernels over the surface elements.
These evaluations effectively take the form of an $N$-body potential: a sum over a set of $N$ positions of products between a kernel and a ``charge'' at each position.
In this case, the locations of the $N$ particles are the centroids of the surface triangularization elements.
A Cartesian particle-cluster treecode is used to compute matrix-vector products and reduce the computational cost of this dense system from $\mathcal{O}(N^2)$ to $\mathcal{O}(N\log N)$ for $N$ points on the discretized molecular surface~\cite{Li2009}.
In particular, to rapidly evaluate the $N$-body potential at the $N$ particle locations, the treecode subdivides the particles into a tree-like hierarchical structure of clusters.
At each location, the potential contribution from nearby particles is computed by direct sum, while for well-separated particle cluster interactions, a Taylor approximation about the center of the cluster is used to evaluate the contribution.
The Taylor coefficients are calculated through recurrence relations.
The resulting linear system is then solved with GMRES iteration~\cite{Saad1986}.

Because the integral equations are defined on the molecular boundary, the singular charges are handled analytically and do not introduce the same issues present in grid-based schemes.
The integral equations also rigorously enforce the interface conditions on the surface, and the boundary condition at infinity is exactly satisfied.
Thus, the boundary integral formulation can potentially be superior to other methods for investigating electrostatic potential on the boundary.

\subsubsection{Boundary element calculation configuration}
APBS users can invoke TABI-PB with the \keyword{bem-manual} flag in the \keyword{ELEC} section of the input file.
Major options include:
\begin{itemize}
	\item \keyword{tree\_order} \param{order}:  An integer indicating the order of the Taylor expansion for determining treecode coefficients.
	Higher values of \param{order} will result in a more accurate -- but more expensive --  calculations.
	A typical choice for this parameter is 3.
	\item \keyword{tree\_n0} \param{number}:  The maximum number of particles allowable in a leaf of the treecode (clusters in the last level of the tree).
	A typical choice for this parameter is 500.
	\item \keyword{mac} \param{criterion}:  Multipole acceptance criterion specifies the distance ratio at which the Taylor expansion is used.
	In general, a higher value of \param{criterion} will result in a more accurate but more expensive computation; while a lower value causes more direct summations and forces the particle-cluster interaction to descend to a finer cluster level. 
	A typical choice for this parameter is 0.8.
	\item \keyword{mesh} \param{flag}:  The software used to mesh the molecular surface; 0~=~MSMS, 1~=~Nano\-Shaper's SES implementation, and 2~=~NanoShaper's Skin implementation.
	See Figure~\ref{fig:surface-mesher} for an example of surface meshes.
	\item \keyword{outdata} \param{flag}: Type of output data file generated; 0~=~APBS OpenDX format~\cite{OpenDX} and 1~=~ParaView format~\cite{ParaView}.
\end{itemize}
Additional information about parameter settings is provided via the APBS website~\cite{APBSweb}. 
TABI-PB produces output including the potential and normal derivative of potential for every element and vertex of the triangularization, as well as the electrostatic solvation energy.
Examples of electrostatic surface potential on the protein 1a63 are shown in Figure~\ref{fig:tabi-mg-comparison} by using MSMS and NanoShaper.

\subsection{Analytical and semi-analytical method implementations} \label{app:pbam}
This appendix provides additional information about the analytical and semi-analytic methods \cite{Felberg2017, Lotan2006} introduced in Section~\ref{sec:pbam}.

\subsubsection{Analytical method (PB-AM) background}
The solution to the PB-AM model is represented as a system of linear equations:
\begin{equation} \label{eq:pbam_solve}
	{A} = \Gamma \cdot (\Delta \cdot T \cdot {A} + {E}),
\end{equation}
where \({A}\) represents a vector of the effective multipole expansion of the charge distributions of each molecule, \({E}\) is a vector of the fixed charge distribution of all molecules, \(\Gamma\) is a dielectric boundary-crossing operator, \(\Delta\) is a cavity polarization operator, and \(T\) is an operator that transforms the multipole expansion from the global (lab) coordinates to a local coordinate frame. 
The unknown \({A}\) determined using the Gauss-Seidel iterative method and can then be used to compute physical properties such as interaction energies, forces, and torques. 
The interaction energy for molecule \(i\), ($\Omega^{(i)}$) is given in Eq.~\ref{eq:pbam_energy}.
\begin{equation}\label{eq:pbam_energy}
\Omega^{(i)}=\frac{1}{\epsilon_s} \left \langle \sum_{j \ne i}^N  T \cdot A^{(j) } ,  A^{(i) } \right \rangle 
\end{equation}
where $\epsilon_s$ is the dielectric constant of the solvent and $\langle  M, N \rangle$ denotes the inner product.
When energy is computed, forces follow as:
\begin{equation}\label{eq:pbam_force}
	\textbf{F}^{(i)} = \nabla_i \Omega^{(i)}=\frac{1}{\epsilon_s} [ \langle \nabla_i \,T \cdot A^{(i) } ,  A^{(i) } \rangle +  \langle T \cdot A^{(i) } ,   \nabla_i \, A^{(i) } \rangle ]
\end{equation}
By definition, the torque on a charge in the molecule is the cross product of its position relative to the center of mass of the molecule with the force it experiences.
The total torque on the molecule is a linear combination of the torque on all charges of the molecule, as illustrated in Eq.~\ref{eq:pbamtor}.
\begin{equation}\label{eq:pbamtor}
	\tau^{(i)} =  \frac{1}{\epsilon_s}\left [  ^xH^{(i)}, ^yH^{(i)}, ^zH^{(i)} \right] \times   \left [  \nabla_i L^{(i)} \right ]
\end{equation} 
where $^\alpha H_{n,m}^{(i)}  = \sum_{j=1}^{M_i} \alpha_{j}^{(i)} \gamma_n^{(i)} q_j^{(i)} (\rho_j^{(i)})^n Y_{n,m} (\vartheta_j^{(i)}, \varphi_j^{(i)}) \, , \, \alpha = x , y, z$, is a coefficient vector for each of the charges in the molecule, $M_i$ is the number of charges in molecule $i$, $q_J^{(i)}$ is the magnitude of the $j^{th}$ charge, and $p_j^{(i)}=\Big[\rho_j^{(i)},\vartheta_j^{(i)},\varphi_j^{(i)}\Big]$ is its position in spherical coordinates.
For more details on the PB-AM derivation, see Lotan and Head-Gordon~\cite{Lotan2006}.

\subsubsection{Semi-analytical method (PM-SAM) background}
The derivation details of PB-SAM have been reported  previously~\cite{Yap2010, Yap2013}, with the main points being summarized in this section.
The electrostatic potential ($\phi_r$) of the system at any point $r$ is governed by the linearized form of the PB equation:
\begin{equation}
	-\nabla \cdot \epsilon \nabla \phi + \kappa^2 \phi = \rho,
	\label{eq:pbsam_lpbe}
\end{equation}
where $\kappa$ is the inverse Debye length.
Eq.~\ref{eq:pbsam_lpbe} is a linearization of Eq.~\ref{eqn:pbe} for $\beta q_i \phi \ll 1$.
For the case of spherical cavities, we can solve Eq.~\ref{eq:pbsam_lpbe} by dividing the system into inner sphere and outer sphere regions, and enforcing a set of boundary conditions that stipulate the continuity of the electrostatic potential and the electrostatic field at the surface of each sphere.
The electrostatic potential outside molecule ($I$) is described by:
\begin{equation}
	\phi_{out}^{(i)} ({r})  = \sum_{I=1}^{N_{mol}} \left( 4 \pi \int_{d\Omega^{(I)}} \frac{e^{-\kappa | r - r'|}}{|r - r'|} h^{(I)} (r') dr'  \right)
	\label{eq:pbsam_phi_in}
\end{equation}
where \(h(r)\) is an effective surface charge that can be transformed into the unknown multipole expansion \(H^{(I,k)}\) with inside molecule \textit{I} and sphere \textit{k}.
In a similar manner, the interior potential is given as
\begin{equation}
	\phi_{in}^{(i)} ({r})  = \sum_{\alpha = 1}^{N_C^{(I)}} \frac{1}{|r-r_{\alpha}^{(I)}|} \cdot \frac{q_\alpha^{(I)}} {\epsilon_{in}} + \frac{1}{4\pi} \int_{d\Omega^{(I)}} \frac{1}{|r - r'|} f^{(I)} (r') dr' 
	\label{eq:pbsam_phi_out}
\end{equation}
where \( N_C^{(I)}\) is the number of charges in molecule \textit{I}, \( q_\alpha\) 
is the magnitude of the \(\alpha\)-th charge, \( r_\alpha^{(I)} = \Big[\rho_{\alpha}^{(I)} , \theta_{\alpha}^{(I)} , \phi_{\alpha}^{(I)} \Big]\) is its position in spherical 
coordinates, and \(f(r)\) is a reactive surface charge that can
be transformed into the unknown multipole expansion \(F^{(I,k)}\).
The reactive multipole and the effective multipole, \(H^{(I,k)}\), are given as:
\begin{align}
	F_{n,m}^{(I,k)} &\equiv  \frac{1}{4\pi} \int_{d\Omega^{(I,k)}} f^{(I,k)}(r') \left (  \frac{ a^{(I,k)}}  {r'} \right ) ^{n+1} \overline{ Y^{(I,k)}_{n,m}} (\theta' , \phi') dr'
	\label{eq:fmat} \\
	H_{n,m}^{(I,k)} &\equiv  \frac{1}{4\pi} \int_{d\Omega^{(I,k)}} h^{(I,k)}(r') \left (  \frac{ r'}{a^{(I,k)}} \right ) ^{n} \hat{i}_n(\kappa r') \overline{ Y^{(I,k)}_{n,m}} (\theta' , \phi') dr'
	\label{eq:hmat}
\end{align}
where \(Y_{n,m}\) is the spherical harmonics, $\overline{Y_{n,m}}$ is the complete conjugate, and \(a^{(I,k)}\) is the radius of sphere \textit{k} of molecule \textit{I}.
These multipole expansions can be iteratively solved using:
\begin{align}
	F_{n,m}^{(I,k)} &= \langle I_{E, n,m} ^{(I,k)}, WF^{(I,k)} \rangle \label{eq:fmat_it} \\
	H_{n,m}^{(I,k)} &= \langle I_{E, n,m} ^{(I,k)}, WH^{(I,k)} \rangle
	\label{eq:hmat_it}
\end{align}
where $WF^{(I,k)}$ and $WH^{(I,k)}$ are scaled multipoles computed from fixed charges and polarization charges from other spheres.
\(I_{E,n,m}^{(I,k)}\) is a matrix of the surface integrals over the exposed surface:
\begin{equation}
	I_{E, n,m} ^{(I,k)} \equiv \frac{1}{4\pi } \int_{\phi_E} \int_{\theta_E} Y_{l,s}^{(I,k)}  (\theta', \phi') \overline{Y_{n,m}^{(I,k)}} (\theta', \phi') \sin \theta' d\theta' d \phi' 
	\label{eq:imat}
\end{equation}
Using the above formalism, physical properties of the system, such as interaction energy, forces and torques can also be computed.
The interaction energy of each molecule, ($\Omega^{(i)}$), is the product of the molecule's total charge distribution (from fixed and polarization charges) with the potential due to external sources.
This is computed as the inner product between the molecule's multipole expansion, ($H^{(I,k)}$), and the multipole expansions of the other molecules in the system, ($LHN^{(I,k)}$) as follows:
\begin{equation}
	\Omega^{(i)} = \frac{1}{\epsilon_s} \sum_{k}^{N_k^{(I)}}  \langle LHN^{(I,k)}, H^{(I,k)} \rangle
	\label{eq:pbsam_en_interact}
\end{equation}
which allows us to define the force which is computed as the gradient of the interaction energy with respect to the position of the center of molecule \textit{I}:
\begin{equation}
	F^{(I)} = - \nabla \Omega^{(I)} =  -  \frac{1}{\epsilon_s} \sum_{k}^{N_k^{(I)}} f_{I,k} = -  \frac{1}{\epsilon_s} \sum_{k}^{N_k^{(I)}}( \langle \nabla LHN^{(I,k)}, H^{(I,k)} \rangle +  \langle LHN^{(I,k)}, \nabla H^{(I,k)} \rangle)
	\label{eq:pbsam_force}
\end{equation}
As in the analytical PB-AM method, the torque on a charge in the molecule is the cross product of its position relative to the center of mass of the molecule with the force it experiences.
For a charge at position \(P\) about the center of mass \(c^{(I)}\) for molecule \(I\), the torque is given by the cross product of its position \(r_P^{(I,k)}\) with respect to the center of mass and the force on that charge $f_P$.
We can re-express \(r_P^{(I,k)}\) as the sum of vectors from the center of molecule $I$ to the center of sphere \(k\) (\(c^{(I,k)}\)) and from the center of sphere \(k\) to point $P$ \((r_P^{(I,k)})\).
The total torque on molecule \(I\) is then given by Eq.~\ref{eq:pbsam_torque}.
\begin{equation}
	\tau^{(I)} =  \sum_{k}^{N_k^{(I)}} c^{(I,k)} \times   f_{I,k} + \sum_{k}^{N_k^{(I)}}  \sum_{P\in k} r_P^{(I,k)} \times f_P 
	\label{eq:pbsam_torque}
\end{equation}
where \(f_{I,k}\) is given in Eq.~\ref{eq:pbsam_force} and
\begin{equation}
	f_P = -  \frac{1}{\epsilon_s} \sum_{k}^{N_k^{(I)}}( \langle \nabla_I LHN^{(I,k)}, H_P^{(I,k)} \rangle +  \langle LHN^{(I,k)}, \nabla_I H_P^{(I,k)} \rangle)
	\label{eq:pbsam_torque_fp}
\end{equation}
where
\begin{align}
	H_{P,n,m}^{(I,k)} &= h(\theta_p, \phi_p) Y_{n,m}^{(I,k)} (\theta_p, \phi_p) 
	\label{eq:pbsam_hp} \\
	\nabla_j H_{P,\alpha, n,m}^{(I,k)} &= \left [ \nabla_j \, h(\theta_p, \phi_p) \right ]_{\alpha}   Y_{n,m}^{(I,k)} (\theta_p, \phi_p) 
	\label{eq:pbsam_derv_hp}
\end{align}
where \(\alpha = x , y, z\).
For the derivation of the PB-SAM solver please see the previous publications~\cite{Yap2010, Yap2013} .

\subsubsection{PB-AM and PB-SAM configuration in APBS}
PB-AM and PB-SAM have been fully integrated into APBS, and is invoked using the keyword \keyword{pbam-auto} or \keyword{pbsam-auto} in the \keyword{ELEC} section of an APBS input file.
Major options include:
\begin{itemize}
	\item \keyword{runname} \param{name}: Desired name to be used for outputs of each run.
	\item \keyword{pbc} \param{length}:  Size of the periodic simulation/calculation domain.
	\item \keyword{runtype dynamics}:  Perform a Brownian Dynamics simulation.
	\item \keyword{ntraj} \param{number}:  Number of Brownian Dynamics simulations to run. 
	\item \keyword{term} \param{type} \param{value} \param{mol}:  Allows the user to indicate conditions for the termination of each BD trajectory.
	The following values of \param{type} are allowed:
	\begin{itemize}
		\item \keyword{time} \param{time}: A limit on the total simulation time.
		\item \keyword{x} or \keyword{y} or \keyword{y} or \keyword{z} or \keyword{r} and \param{$>=$} or \param{$<=$}: Represents the approach of two molecules to a certain distance \keyword{r} or certain region of space given by \keyword{x} or \keyword{y} or \keyword{y}.
		The operators $>=$ and $<=$ represent the corresponding inequalities.
	\end{itemize}
	The parameter \param{mol} is the molecular index that this condition applies.  \param{mol} should be 0 for \keyword{time} and for a termination condition of \keyword{x}, \keyword{y} or \keyword{z}, the molecule index that this termination condition applies to.
	\item \keyword{xyz} \param{idx} \param{fpath}:  Molecule index \param{idx} and file path \param{fpath} for the molecule starting configurations. 
	A starting configuration is needed for each molecule and each trajectory.
	Therefore, jf there are $m$ molecules and \keyword{ntraj} \param{n} trajectories, then the input file must contain $m \times n$ \keyword{xyz} entries.
	\item \keyword{tolsp} \param{val}: Modify the coarseness of the molecular description.
	\param{val} is the distance (in \AA) beyond the solvent-excluded surface that the coarse-grained representation extends.
	Increasing values of \param{val} leads to fewer coarse-grained spheres, faster calculation times, but less accurate solutions.
	Typical values for \param{val} are between 1 and 5 \AA.
\end{itemize}
The commands (keywords) not included in this list are used to specify system conditions, such as temperature and salt concentration. These parameters are similar to those found in the \keyword{ELEC} section of a usual APBS run and are documented on the Contributions portion of the APBS website~\cite{APBSweb}. 
Additional information about parameter settings is provided via the APBS website~\cite{APBSweb}.
\revision{Examples of the electrostatic potentials produced from PB-AM and PB-SAM are shown in Figure~\ref{fig:pbsam-results}.}

\printbibliography

\end{document}